%% file: Final-arXiv.tex
\documentclass[lettersize,journal]{IEEEtran}
\usepackage{amsmath, amssymb, amsfonts, bm, mathtools}

\usepackage{mathdots}
\usepackage{yhmath}
\usepackage{cancel}
\usepackage{siunitx}
\usepackage{extarrows}
\usepackage{algorithm}
\usepackage{algpseudocode}
\usepackage{graphicx}
\usepackage{subcaption}
\usepackage{caption}
\usepackage{array}
\usepackage{multirow}
\usepackage{tabularx}
\usepackage{booktabs}
\usepackage{tikz}
\usetikzlibrary{shapes, shadows, arrows, fadings, patterns, shadows.blur}
\usepackage{cite}
\usepackage{url}

\usepackage[nolist]{acronym}
\usepackage{xcolor}
\usepackage{soul}
\usepackage[top=0.8in, left=0.65in, right=0.65in, bottom=1.1in]{geometry}

\usepackage[utf8]{inputenc}
\usepackage{subcaption}  

\begin{document}
\pagenumbering{arabic}

\title{{Adaptive Machine Learning Framework for UAV Trajectory Optimization in O-RAN}}

\author{\IEEEauthorblockN{Chenrui Sun \IEEEmembership{Student Member,~IEEE}, Swarna Bindu Chetty \IEEEmembership{ Member,~IEEE}, Gianluca Fontanesi \IEEEmembership{ Member,~IEEE}, Mahnaz Arvaneh \IEEEmembership{ Member,~IEEE}, Walid Saad \IEEEmembership{Fellow,~IEEE} and Hamed Ahmadi \IEEEmembership{Senior Member,~IEEE}}
\\

\thanks{C. Sun, S.B. Chetty and H. Ahmadi are with the school of Physics, Engineering and Technology, University of York, United Kingdom. G. Fontanesi is with Radio Systems Research, Nokia Bell Labs, Stuttgart, Germany. M. Arvaneh is with the School of Electrical and Electronic Engineering, University of Sheffield, United Kingdom. W. Saad is with the Bradley Department of Electrical and Computer
Engineering, Virginia Tech, VA, 24061, USA}

}
\markboth{Accepted for publication at IEEE TVT 2026}%
{Shell \MakeLowercase{\textit{et al.}}: A Sample Article Using IEEEtran.cls for IEEE Journals}

\maketitle

\begin{abstract}

The deployment of unmanned aerial vehicles (UAV) as open radio units (O-RUs) in 6G cellular systems presents a promising opportunity to achieve scalable and adaptive network coverage. However, optimizing UAV trajectories in dynamic and unfamiliar environments remains a critical challenge, particularly due to the need for extensive retraining in each new scenario. In this paper, we introduce a novel UAV trajectory optimization framework that integrates enhanced continual transfer learning within the \ac{O-RAN} architecture. The proposed system maintains a library of pre-trained models and employs a model selection mechanism to identify and transfer knowledge from the most relevant environments, minimizing adaptation time and improving efficiency. When no sufficiently similar model is available, a fallback model empowered by continuous refinements ensures baseline performance. The framework leverages real-world city maps and ray tracing techniques to enhance learning reliability and improve trajectory planning. Simulation results demonstrate that the proposed model selection-based transfer learning approach reduces convergence time by 44\% to 56\% compared to retraining from scratch, and up to 40\% compared to traditional transfer learning without model selection.

\end{abstract}


\section{Introduction}

\ac{6G} wireless cellular systems are expected to provide ultra-low latency, massive connectivity, and unprecedented adaptability to diverse scenarios. The leap towards \ac{6G} requires a rethinking of traditional infrastructure deployment to meet the demands of dynamic environments such as disaster zones, dense urban areas, and remote regions. To achieve this, open radio access network (O-RAN) architecture is expected to be the cornerstone of \ac{6G} innovation \cite{ahmadi2025towards}. \ac{O-RAN} introduces an open and intelligent architecture that disaggregates traditional RAN functions and standardizes interfaces, enabling multi-vendor interoperability and enhanced control through centralized intelligence. Within this framework, \ac{UAV} present a new opportunity to operate as agile and reconfigurable \acp{O-RU}, dynamically extending coverage in the \ac{O-RAN} architecture \cite{chenrui}. For instance, \ac{UAV}-based \acp{O-RU} can provide connectivity in disaster affected regions, enhance capacity at crowded events, or deliver network access to remote communities. Their agility and mobility make \acp{UAV} an essential component of the \ac{O-RAN} framework, ensuring rapid and scalable deployment of \ac{6G} networks. 

Due to limited capacity of \ac{UAV} batteries, minimizing flight distance is crucial to extending mission duration and ensuring efficient coverage of objectives. This requires better optimization of the \ac{UAV}’s trajectory \cite{mozaffari2019tutorial}. At the same time, maintaining a good connection with the ground \acp{O-DU} is critical to support communication needs during flight. To meet the dual challenges of energy efficiency and communication quality, trajectory planning must consider environmental dynamics, user locations, and signal quality. Traditional optimization methods are insufficient in such complex and time-varying scenarios. In contrast, \ac{ML} techniques, particularly \ac{RL}, have shown promising results. \ac{RL} enables \acp{UAV} to learn adaptive trajectory policies through interaction with the environment, balancing exploration and exploitation to discover efficient paths \cite{9625502}.
 \begin{figure}[t]
    \centering
    \includegraphics[width=0.9\columnwidth]{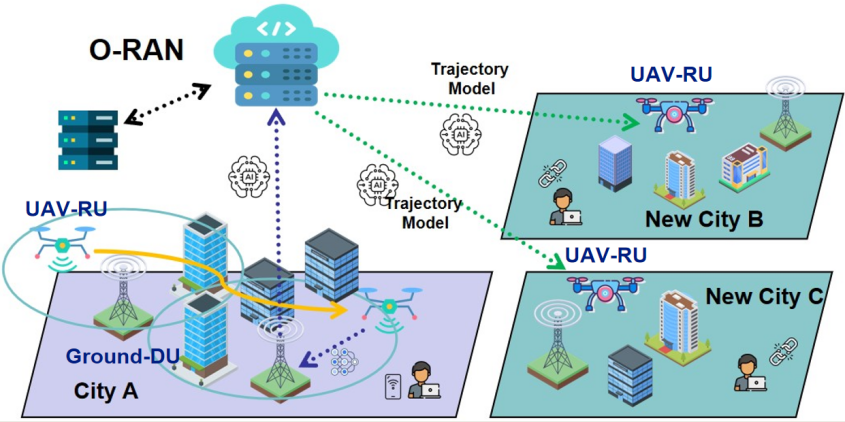}
    \caption{Environmentally Adaptive UAV Trajectory Optimization with O-RAN System}
    \label{fig:showcase}
\end{figure}

However, a significant limitation of \ac{RL} is its poor adaptability across environments. Once deployed in a new scenario with different urban structures, terrain, or communication characteristics, an \ac{RL} model typically requires retraining from scratch. This retraining process is computationally expensive and time-consuming, which hinders real-time \ac{UAV} operations and scalability across diverse settings. To overcome this, \ac{TL} has been proposed as a way to reuse knowledge from previously trained models to accelerate learning in new environments. \ac{TL} can significantly reduce retraining time and improve initial performance. However, the practical deployment of \ac{TL} in \ac{UAV} trajectory planning still faces several challenges. Specifically, it is unclear how to evaluate the applicability of the learning path to a specific target environment, how to identify the most suitable source model from the pre-trained library, and under what conditions knowledge transfer leads to performance improvement rather than negative transfer. Addressing these issues is critical to ensure that \ac{TL} not only accelerates convergence speed but also ensures robust and reliable operation of \acp{UAV} in heterogeneous urban scenarios.

This work focuses on these critical questions. We design a model selection framework that enhances the reliability and performance of \ac{TL} for \ac{UAV} trajectory optimization. Our approach not only selects suitable models based on environmental similarity but also integrates this mechanism into the \ac{O-RAN} platform, enabling real-time adaptation, continuous learning, and practical deployment. By combining \ac{TL} with enhanced continuous learning and embedding the solution into the \ac{O-RAN} architecture. This is able to address the limitations of previous \ac{ML}-based trajectory systems.
\subsection{Prior Works}
Recent advances in \ac{ML} have significantly enhanced \ac{UAV} trajectory optimization, particularly in dynamic and complex wireless environments. \ac{RL} has been widely adopted to enable \acp{UAV} to autonomously learn optimal paths through interaction with their surroundings. Early studies such as \cite{pham2018autonomous} and \cite{yin2019intelligent} showed the feasibility of \ac{RL} for autonomous \ac{UAV} navigation and trajectory design. \ac{DRL} techniques, including \ac{DDPG} and \ac{DDQN}, have further improved trajectory learning under uncertainty and mobility constraints, as shown in \cite{bouhamed2020autonomous, ding20203d, moon2021deep}. Energy efficiency has emerged as a parallel optimization goal. In \cite{zhu2021uav} and \cite{ hu2023reinforcement},  the authors used \ac{DRL} and knowledge transfer to reduce \ac{UAV} energy consumption while maintaining connectivity. Similarly, the work in \cite{ding20203d} incorporated fairness and frequency allocation alongside energy-aware trajectory planning. Moreover, in \cite{yang2021efficient}, the authors applied heuristic learning to meet connectivity constraints in cellular-connected \acp{UAV}, while the work in \cite{bhandarkar2021optimal} combined \ac{RL} with greedy algorithms to balance performance and learning speed in mobile base station applications. Additionally, model-based and optimization-driven methods remain relevant for scenarios requiring 3D placement and \ac{UAV}-assisted network design \cite{lakew2020three}. Connectivity-aware planning using aerial coverage maps was also proposed in \cite{yang2019connectivity} to proactively guide \ac{UAV} paths based on prior knowledge. 

More recently, \ac{TL} has been employed to improve generalization and reduce training time across environments. In \cite{zhang2020trajectory} and \cite{fontanesi2022transfer}, the authors demonstrated that transferring models between environments significantly accelerates adaptation and improves reliability, especially under outage constraints. In parallel, the work in \cite{chen2020knowledge} explored inter-\ac{UAV} knowledge transfer for trajectory tracking, enabling faster learning in new missions. Recent studies have also investigated meta-learning for rapid adaptation by learning shared knowledge across tasks, and many of these approaches rely on repeated task-level training, whereas transfer learning focuses on reusing pre-trained models to reduce retraining cost in deployment scenarios \cite{eldeeb2025multi}.

Recent research has also explored distributed artificial intelligence and multi-modal learning frameworks for intelligent control in future wireless and UAV-assisted networks. In particular, \cite{DistributedAI_SAGS_2025} investigates distributed AI-based secure communications across space-air-ground-sea integrated networks, while \cite{DistributedFM_6G_2024} introduces distributed foundation models for multi-modal learning in 6G networks under O-RAN architectures. These studies indicate that UAV trajectory optimization can benefit from multi-modal inputs, such as radio environment maps, sensing data, and network state information, with coordination supported by RAN Intelligent Controllers. However, most existing studies focus on single-environment learning and lack mechanisms for continuous adaptation in real-world heterogeneous environments. In addition, there is a lack of systematic analysis on under what conditions \ac{UAV} trajectory models are suitable for \ac{TL} and which types of source environments can produce the most effective transfer performance. In Table~\ref{tab:review}, we review relevant studies on \ac{UAV} trajectory optimization using \ac{ML} and highlight the distinctions of our contribution.

\begin{table*}[ht]
\centering
\caption{Summary of Machine Learning-Based UAV Trajectory Studies}
\label{tab:review}
\begin{tabularx}{\textwidth}{l r l l l X}
\toprule
\textbf{Reference} & \textbf{Year} & \textbf{ML Method} & \textbf{Focus} & \textbf{Scenario} & \textbf{Application} \\
\midrule
\cite{yang2019connectivity} & 2019 & Mapping with RL & Trajectory & UAV & Connectivity-aware path \\
\cite{lakew2020three} & 2020 & Optimization with RL & Placement \& Trajectory & UAV-assisted Network & 3D optimization \\
\cite{pham2018autonomous} & 2018 & RL & Trajectory & Single UAV & Autonomous navigation \\
\cite{yin2019intelligent} & 2019 & RL & Trajectory & Single UAV & Urban coverage \\
\cite{ebrahimi2020autonomous} & 2020 & RL & Trajectory & Single UAV & Object localization \\
\cite{bouhamed2020autonomous} & 2020 & DDPG & Trajectory & Single UAV & Autonomous navigation \\
\cite{ding20203d} & 2020 & DRL & Energy \& Fairness & Multi-UAV & 3D trajectory + frequency allocation \\
\cite{zhu2021uav} & 2021 & DRL & Energy & WSNs & Energy minimization \\
\cite{moon2021deep} & 2021 & DRL & Trajectory & Multi-UAV & Target tracking \\
\cite{bhandarkar2021optimal} & 2021 & RL + Greedy & Trajectory & UAV-MBS & Optimal learning \\
\cite{hu2023reinforcement} & 2023 & RL + Transfer & Energy & UAV-BS & Efficiency improvement \\
\cite{ding2023ddqn} & 2023 & DDQN & Trajectory + Resource & UAV-MEC & Secure communications \\
\cite{yang2021efficient} & 2021 & Heuristic + ML & Trajectory & Cellular UAV & Connectivity constraints \\
\cite{chen2020knowledge} & 2020 & Transfer Learning & Trajectory Tracking & Multi-UAV & Cross-UAV adaptation \\
\cite{zhang2020trajectory} & 2020 & Transfer Learning & Trajectory & UAV & Emergency communication \\
\cite{fontanesi2022transfer} & 2022 & Transfer Learning & Trajectory & UAV & Outage-aware design \\
This work & 2025 & DDQN+TL+CFL  & Variable Target Trajectory & Adaptive UAV with Muti-Cities& Training time and Energy saving\\
\bottomrule
\end{tabularx}
\end{table*}

\subsection{Contributions}
The main contribution of this paper is introducing a flexible and adaptive framework for deploying \acp{UAV} as \ac{O-RU} in \ac{6G} networks, addressing key challenges in trajectory optimization and adaptability in dynamic environments. In summary, our key contributions include:

\begin{itemize}
     \item We develop a model selection mechanism that selects the most appropriate pre-trained model from a library of trajectory policies based on an environment similarity metric. A fallback general model \( M_G \) is employed when no sufficiently similar model exists. This approach significantly reduces the retraining overhead and further improves transfer efficiency.

      \item We created a fallback model \( M_G \) that is continuously updated based on new environments, \( M_G \) initially trained on synthetic simulation environments, it will update using \ac{CL} as \acp{UAV} encounter new environments. This enables the system to progressively accumulate knowledge across diverse urban scenarios.

       \item We develop a integrated \ac{UAV} trajectory system within \ac{O-RAN} structure. Model selection and training are integrated through xApps and rApps within the Near-RT and Non-RT RICs, enabling real-time inference and continuous learning.
       
      \item We validate our framework using ray-traced RSSI maps from real cities including York, Beijing, and Ottawa. The simulation settings replicate realistic propagation conditions such as diffraction and multipath, demonstrating the framework’s effectiveness in practical scenarios
\end{itemize} 

\section{UAV Trajectory System Model and Problem formulation}

In this section, we present the overall system architecture, including trajectory model, and problem formulation. The main goal of this work is to enhance the adaptability and training speed of \acp{UAV} trajectory in different environments (cities) by strengthening the effectiveness of \ac{TL}. Particularly, we aim to improve the selection of source models such that transferred knowledge is more relevant and useful, thereby accelerating training and improving performance. To contextualize the \ac{UAV} trajectory planning task, Section ~\ref{tra} details our system model and environment settings, focusing on energy constraints and communication requirements with ground \ac{O-DU}. To increase the flexibility and realism of the trajectory planning, we also simulate different target destinations within each city. Building upon this foundation, Section ~\ref{pro} first lists the formulation of the trajectory problem and then defines the main problem to be solved in this study, with a special focus on how to improve the \ac{TL} performance of adaptive \ac{UAV} trajectory optimization.

\begin{figure*}[ht]
\centering
\begin{subfigure}{0.3\textwidth}
    \centering
    \includegraphics[width=\linewidth]{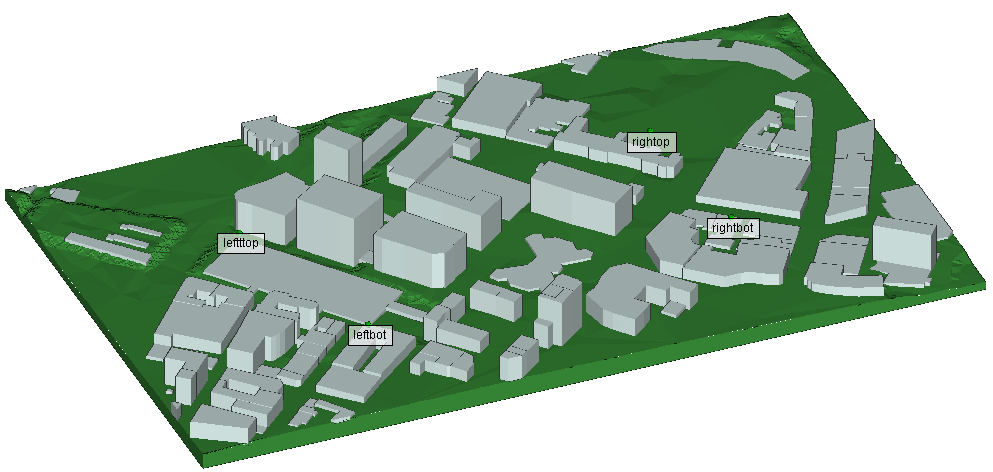}
    \caption{Birmingham}
\end{subfigure}
\hfill
\begin{subfigure}{0.3\textwidth}
    \centering
    \includegraphics[width=\linewidth]{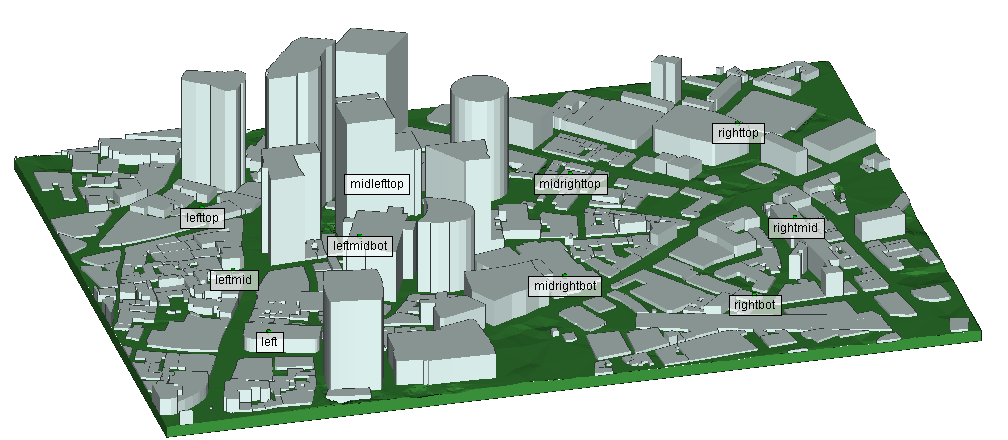}
    \caption{London}
\end{subfigure}
\hfill
\begin{subfigure}{0.3\textwidth}
    \centering
    \includegraphics[width=\linewidth]{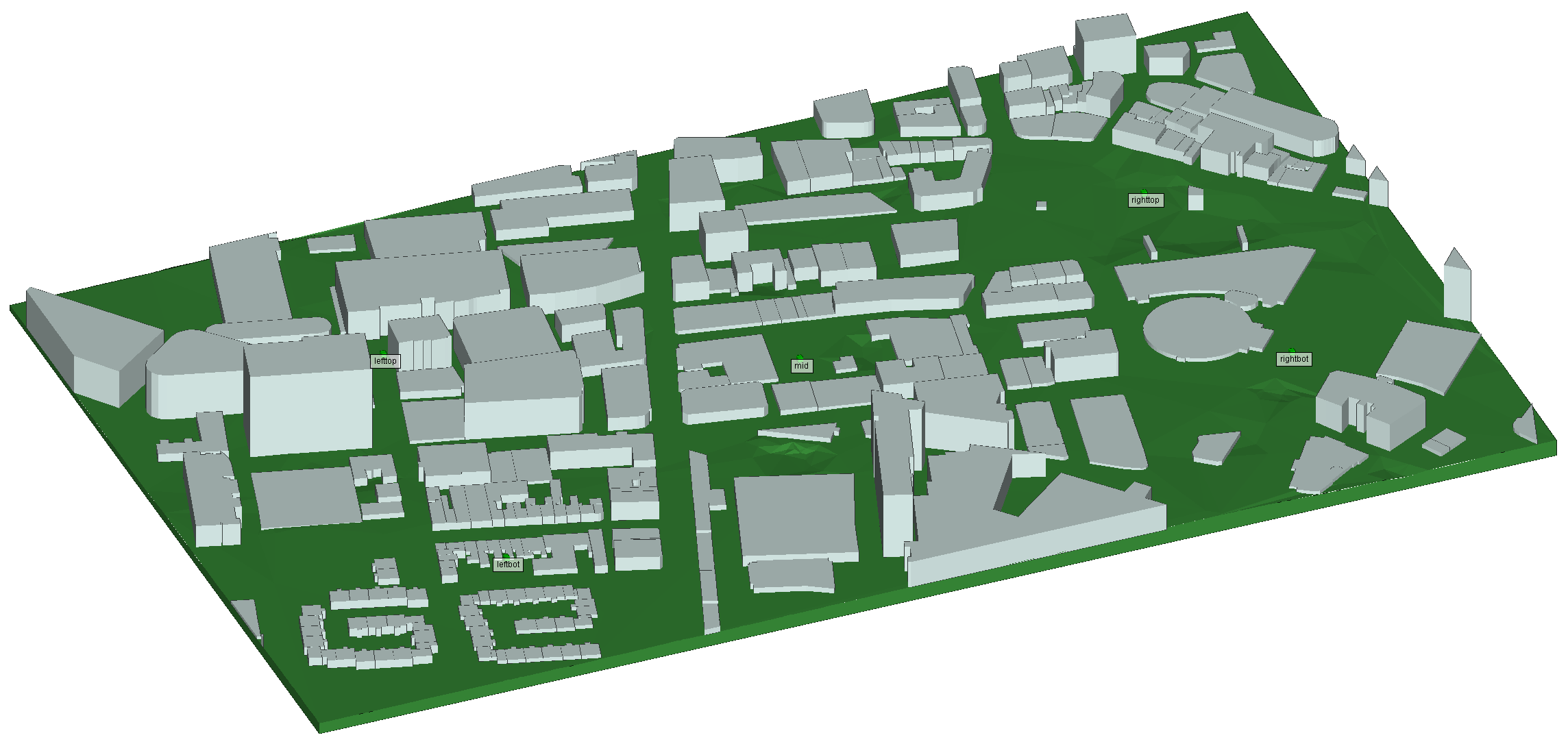}
    \caption{Manchester}
\end{subfigure}

\begin{subfigure}{0.3\textwidth}
    \centering
    \includegraphics[width=\linewidth]{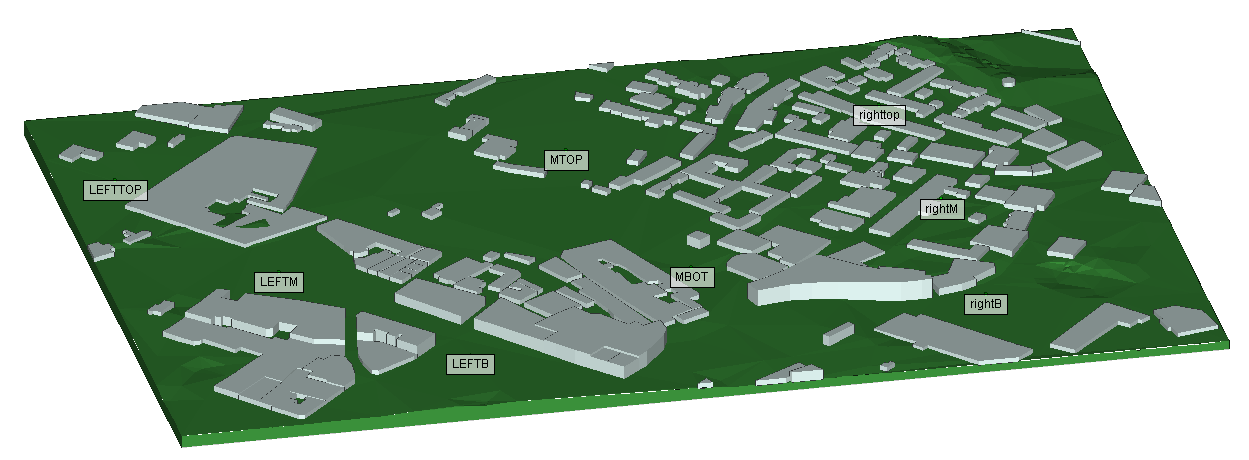}
    \caption{York}
\end{subfigure}
\hfill
\begin{subfigure}{0.3\textwidth}
    \centering
    \includegraphics[width=\linewidth]{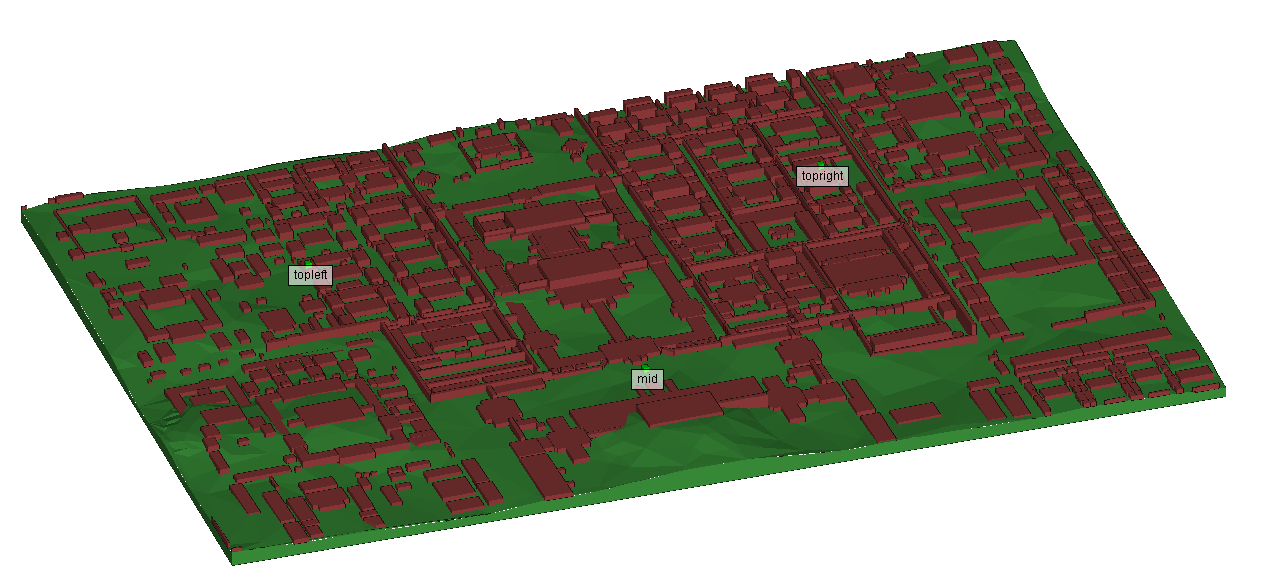}
    \caption{Beijing}
\end{subfigure}
\hfill
\begin{subfigure}{0.3\textwidth}
    \centering
    \includegraphics[width=\linewidth]{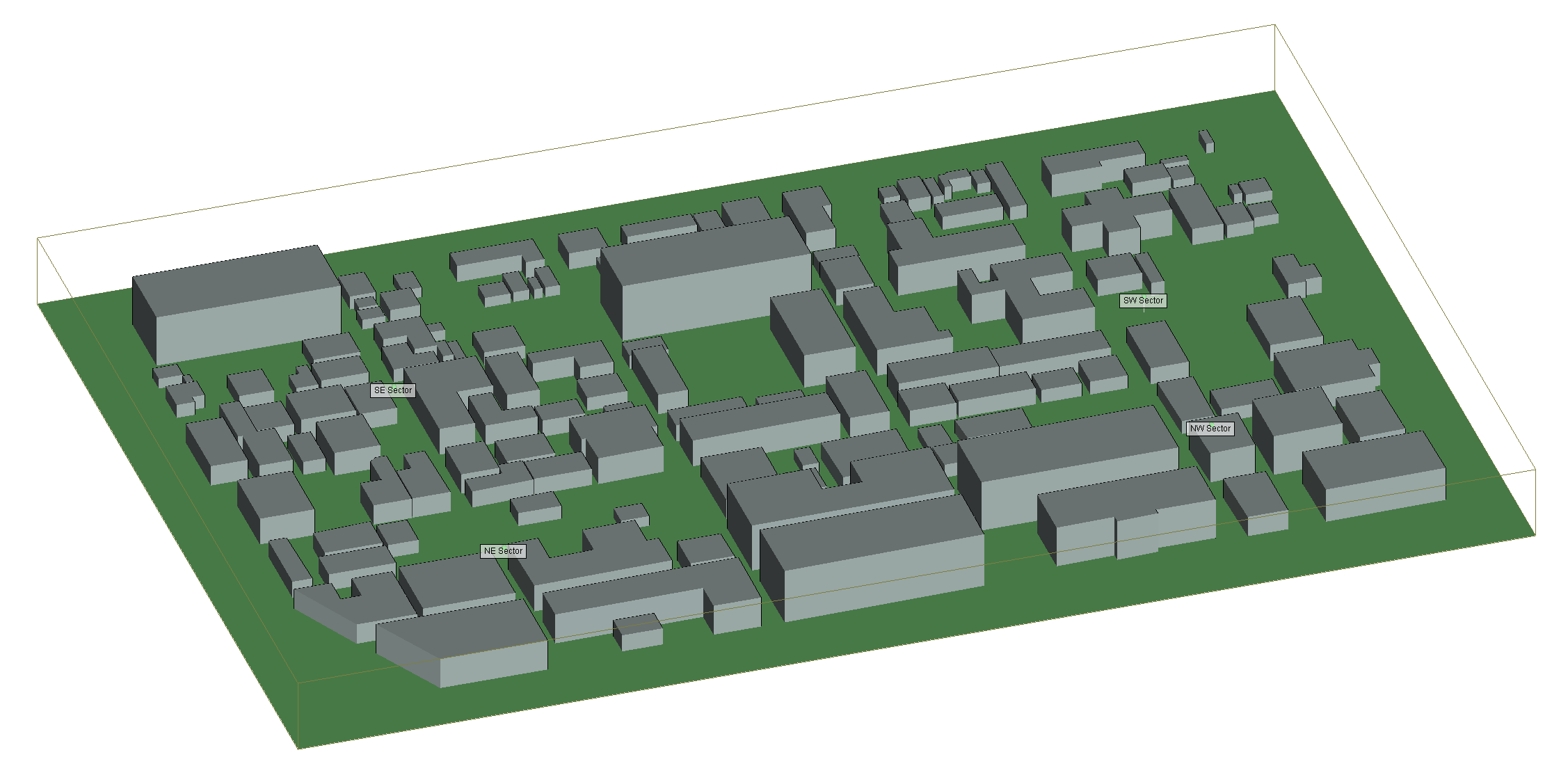}
    \caption{Ottawa}
\end{subfigure}
\caption{3D maps of different urban environments generated by wireless insite, including different city types (urban areas, skyscrapers, classical buildings, residential areas, etc.). Different materials are used for buildings and land according to the actual situation. }
\label{fig:maps}
\end{figure*}


\subsection{UAV Trajectory System Model}\label{tra}

We consider a single \ac{UAV}-aided cellular network, where the \ac{UAV} serves as a mobile \ac{O-RU}, autonomously flying to its destination across various urban environments (cities). These urban environments, each corresponding to a different real-world city map segment (as shown in Fig.~\ref{fig:maps}), are collectively defined as the space of cities \( \mathcal{E} = \{ \mathcal{E}_1, \mathcal{E}_2, \ldots \} \). When the \ac{UAV} encounters a new urban scenario, it is referred to as \( \mathcal{E}_{\text{new}} \). Each cities presents unique characteristics, including variations in building distribution, density, height, street layout, and so on, all of which influence communication requirements and signal propagation.

Ground \acp{BS} consists of \ac{O-DU} which connect to \ac{O-CU} and are deployed at predefined positions within each environment. It also hosts the O-RAN architecture, which will be introduced in Section~\ref{Oran}. The \ac{UAV} must maintain robust connectivity to the \ac{O-DU} which is providing the fronthaul link while navigating through the urban environment. The \ac{UAV} is initially deployed at position \( q_0 = (x_{\text{UAV}}, y_{\text{UAV}}, h_{\text{UAV}}) \in \mathcal{A}_{\text{launch}} \subset \mathbb{R}^3 \), where \( \mathcal{A}_{\text{launch}} \) is the designated launch area. Users are modeled within a defined target region \( \mathcal{A}_{\text{target}} \subset \mathbb{R}^3 \), where \( U \) users are spatially distributed and the distribution of users changes over time. As the user distribution changes, the \ac{UAV} will have a new target position \(q_t\). The \ac{UAV} flies at a constant velocity \( V = V_{\max} \) and must satisfy altitude constraints:
\(
h_{\text{min}} \leq h_{\text{UAV}} \leq h_{\text{max}}
\).
The \ac{UAV}'s trajectory is represented as a sequence of 3D positions:
\[
\mathbf{q} = \{ q(n) = (x_n, y_n, h_n) \mid n = 0, 1, \dots, K \},
\]
where \( K \) is the total number of movement steps.

We use two types of channel models: one for training in simulation environments and one for realistic evaluations using ray tracing. In simulation environments, we adopt a simplified analytical model consistent with \cite{Sun_wincom24}, which captures large-scale path loss, small-scale fading, and environmental effects such as terrain and building obstructions. The large-scale path loss includes both Line-of-Sight (LoS) and Non-Line-of-Sight (NLoS) components, which adopt the ground-to-air (G2A) version of the 3GPP, respectively, as follows:
\begin{equation}
l_{\text{LoS}}(d) = X_{\text{LoS}} \cdot d^{-\alpha_{\text{LoS}}}, \quad
l_{\text{NLoS}}(d) = X_{\text{NLoS}} \cdot d^{-\alpha_{\text{NLoS}}},
\end{equation}
where \( d \) is the distance between the \ac{UAV} and ground \acp{RU} and DUs, \( \alpha \) is the path loss exponent, and \( X \) is the path loss constant that is defined based on the environmental type.

For realistic evaluation, we adopt a ray-tracing-based channel model using full 3D propagation. The simulation follows the Shooting and Bouncing Ray (SBR) method \cite{wang2022sbr}, accounting for up to \( R \) reflections and \( D \) diffractions per ray path. Let the carrier frequency be \( f_c \), and the system bandwidth be \( B \). The \ac{UAV} and ground receivers are equipped with vertically polarized half-wave dipole antennas, each with gain \( G_t \) and \( G_r \), respectively, and the transmit power is given by \( P_t \). The receiver operates with a noise figure \( N_f \), and the thermal noise power is given by: 
\(N_0 = k T B \cdot 10^{\frac{N_f}{10}}\)
where \( k \) is Boltzmann’s constant and \( T \) is the absolute temperature. The received power at position \( q(n) \) is the sum of all multipath components:
\begin{equation}
P_r(q(n)) = \sum_{l=1}^{L} P_l(q(n)),
\end{equation}
where \( L \) is the number of multipath components and \( P_l(q(n)) \) is the power of the path \( l \). The \ac{SINR} is defined as:
\begin{equation}
\text{SINR}(q(n)) = \frac{P_r(q(n))}{N_0 + I},
\end{equation}
where \( N_0 \) is the thermal-noise power and \( I \) denotes the aggregate interference generated by other terrestrial \ac{O-DU}. Hence, \( I \) represents background network interference and not self-interference from the airborne link. The relevant channel model parameter settings will be introduced in detail in section \ref{env}. In order to quantify and avoid the \ac{UAV} from being disconnected from the ground or entering a weak signal area, an outage event is defined as when the received \ac{SINR} at position \( q(n) \) falls below a threshold \( \phi_{\text{th}} \). We define an indicator function:
\begin{equation}
\beta(q(n)) = 
\begin{cases}
1, & \text{if } \text{SINR}(q(n)) \leq \phi_{\text{th}} \\
0, & \text{otherwise}.
\end{cases}
\end{equation}
The total number of outage events during a trajectory is then computed as:
\begin{equation}
\Gamma = \sum_{n=0}^{K} \beta(q(n))
\end{equation}

\subsection{Problem Formulation}\label{pro}

In order to define what kind of trajectory model we want to make sustainable and adaptable, we first clarify the trajectory problem formulation. The \ac{UAV} trajectory system aims to optimize the flight path from its initial position \( q_0 \in \mathcal{A}_{\text{launch}} \) to a target location \( q_t \in \mathcal{A}_{\text{target}} \), and target location changes dynamically based on the distribution of users~\cite{Sun_wincom24}. The trajectory is defined as \( \mathbf{q} = \{q(n)\}_{n=0}^{K} \), where each \( q(n) = (x_n, y_n, h_n) \in \mathbb{R}^3 \) represents the \ac{UAV}'s position at time step \( n \). The optimization objective for trajectory problem is defined as:

\begin{subequations}\label{eq:ProblemFormulation}
\begin{align}
\mathbf{q}^* &= \arg\min_{\mathbf{q}}\!\Bigl[S(\mathbf{q}) + w\,\Gamma(\mathbf{q})\Bigr], \label{eq:ProblemFormulation_obj}\\[4pt]
\text{s.t.}\quad 
\mathbf{q}(0) &= q_0,\quad \mathbf{q}(K) = q_t,  \label{eq:initial_final}\\
h_{\text{min}} &\le h_n \le h_{\text{max}},
    \label{eq:altitude_constraint}\\
K &\le K_{\max}.
    \label{eq:max_steps}
\end{align}
\end{subequations}

\( S(\mathbf{q}) \) is the number of steps (total travel distance), and \( \Gamma(\mathbf{q}) \) represents the total number of outage events along the trajectory. The weights \( w\)  balance the tradeoff between minimizing travel and maintaining service quality. Constraint~\eqref{eq:initial_final} ensures that the \ac{UAV} begins its trajectory at a known initial position \( \mathbf{q}(0)\) and reaches the dynamically determined target location\(\quad q(K)\). These positions define the start and end points of the mission within the environment. Constraint~\eqref{eq:altitude_constraint} ensures that the \ac{UAV}’s altitude \( h_n \) at each step remains within allowable bounds. These bounds maintain flight safety and compliance with airspace regulations while influencing connectivity due to LoS constraints. Constraint~\eqref{eq:max_steps} imposes an upper limit on the total number of trajectory steps \( K \), which corresponds to the \ac{UAV}’s energy budget or battery capacity. This restricts how far or how long the \ac{UAV} can operate within a given mission. Our prior works such as \cite{sun2025energy} have solved problem \eqref{eq:ProblemFormulation} using \ac{RL} approaches. However, \ac{RL} methods cannot efficiently adapt to new environments, as they require retraining from scratch, which consumes significant resources and time. Instead, there is a need for approaches that enable rapid adaptation and \ac{TL} with minimal retraining overhead.

We defined \( \mathcal{E} \) as the space of urban environments, where each environment \( \mathcal{E}_i \in \mathcal{E} \) is characterized by its ground base station distribution, building distribution, and building height. When the \ac{UAV} enters a new environment \( \mathcal{E}_{\text{new}} \notin \mathcal{E}_{\text{trained}} \), it must adapt its trajectory policy to accommodate this environment's unique spatial and communication characteristics. Since \( \pi \) is the policy used by the \ac{UAV} to generate a trajectory \( \mathbf{q}_\pi \) in different \( \mathcal{E} \).  The goal is to discover a policy \( \pi^* \) that minimizes the time required for adaptation, \( T_{\text{adapt}} \). We define the adaptation time as the number of training steps required for the \ac{UAV} to become functionally operational in \( \mathcal{E}_{\text{new}} \), satisfying a predefined performance threshold formally:
\begin{equation}
T_{\text{adapt}} = \min \left\{ t \mid \mathcal{C}(\mathbf{q}_\pi^t, \mathcal{E}_{\text{new}}) \leq \delta \right\},
\end{equation}
where \( \mathcal{C}(\cdot) \) is a performance cost function, and \( \delta \) is the maximum acceptable cost value reflecting mission requirements such as outage limit and trajectory efficiency. The learning approach \( \Theta \) determines how the policy \( \pi \) is initialized and adapted in \( \mathcal{E}_{\text{new}} \). Our objective for the adaptive problem is to find the optimal learning strategy \( \Theta^* \) that results in the lowest adaptation time:
\begin{equation}
\Theta^* = \arg\min_{\Theta \in \mathcal{H}} T_{\text{adapt}}(\Theta, \mathcal{E}_{\text{new}}),
\end{equation}
where \( \mathcal{H} \) is the space of possible learning approaches (from-scratch training, fallbacks, or transferred models). This abstracts the adaptation problem independently from any specific solution method. It defines a general objective: to identify the most efficient strategy \( \Theta \) for adapting \ac{UAV} trajectory policies to previously unseen environments while maintaining required communication and navigation performance.

\section{DDQN and TL with Model selection} \label{ml}
\subsection{Dueling DDQN with Multi-Step Learning for \ac{UAV} Trajectory Optimization}

To establish a strong baseline for trajectory learning, we employ a Dueling Double Deep Q-Network (Dueling \ac{DDQN}) with multi-step. This \ac{DDQN} framework is particularly well-suited for high-dimensional and dynamic \ac{UAV} environments within the \ac{O-RAN} architecture. The \ac{UAV} trajectory optimization problem is modeled as a \ac{MDP}, defined by the tuple \( (\mathcal{S}, \mathcal{A}, P, R, \gamma) \), where \( \mathcal{S} \) denotes the set of states and \( \mathcal{A} \) the set of discrete \ac{UAV} movement actions. Each state \( s_n \in \mathcal{S} \) corresponds to the \ac{UAV}'s position \( q(n) = (x_n, y_n, h_n) \in \mathbb{R}^3 \) at time step \( n \). The goal is to learn an optimal policy \( \pi^* \) that maximizes the expected cumulative reward:
\begin{equation}
\pi^* = \arg\max_\pi \mathbb{E} \left[ \sum_{n=0}^{K} \gamma^n R(s_n, a_n, s_{n+1}) \right],
\end{equation}
The reward function is defined as:
\begin{equation}
R(s_n, a_n, s_{n+1}) =  - d(s_n, s^*) - w_r \cdot \psi(q(n)) + \mathcal{R}_{\text{goal}}- \mathcal{C}_{\text{out}},
\end{equation}
where: \( d(s_n, s^*) \) is the Euclidean distance from the current \ac{UAV} state \( s_n \) to the dynamic target state \( s^* \), \( \mathcal{C}_{\text{out}} \) penalizes the \ac{UAV} for moving outside defined operational boundaries, \( \mathcal{R}_{\text{goal}} \) is a positive reward granted when the \ac{UAV} reaches the dynamically determined target location. The weights \( w_r \)  control the tradeoff between minimizing travel distance and avoiding low-quality signal regions. \( \psi(q(n)) \) is a continuous penalty function based on the outage probability at position \( q(n) \), defined as:
\begin{equation}
\psi(q(n)) = P_{\text{out}}(q(n)) = \Pr\left( \text{SINR}(q(n)) \leq \phi_{\text{th}} \right).
\end{equation}
This represents the probability that the received \ac{SINR} at position \( q(n) \) falls below a threshold \( \phi_{\text{th}} \), indicating poor signal quality. Values of \( \psi(q(n)) \) range from 0 to 1, enabling more nuanced and stable learning behavior. The dueling architecture enhances learning efficiency by independently estimating the value of a state and the advantages of actions, while the multi-step return improves long-term trajectory planning. This baseline has been proven stable and generalizable trajectory learning~\cite{9625502,Sun_wincom24} and serves as a benchmark for evaluating \ac{TL} and \ac{CL} model updates in complex urban scenarios. The system-level parameters, UAV operating constraints, and ray-tracing settings used in the trajectory optimization model are summarized in Table~\ref{tab:parameters}.

\subsection{Advanced \ac{TL} for \ac{UAV} Adaptation in New Environments}

Although \ac{RL} has proven effective for \ac{UAV} trajectory optimization, it suffers from a critical limitation: the learned policy is highly environment specific. When the \ac{UAV} enters a new urban scenario with different building layouts, the \ac{DDQN} agent must be retrained from scratch to adapt. This retraining process is not only time-consuming but also computationally expensive, making it impractical for real-time deployment in dynamic or large-scale settings.

To address this, we apply \ac{TL}, which mitigates the cost of retraining by leveraging previously learned policies from other environments. In \ac{TL}, the \ac{UAV} begins its learning process in a new environment not from random initialization but from a pre-trained policy \( \pi_{\text{source}} \), thus accelerating convergence and improving early-stage performance. The transferred policy is then fine-tuned to the new environment:
\begin{equation}
\pi_{\text{new}}^{(0)} = \pi_{\text{source}}, \quad 
\pi_{\text{new}}^{(k+1)} = \pi_{\text{new}}^{(k)} - \alpha \cdot \nabla_{\pi} \mathcal{L}(\pi_{\text{new}}^{(k)}),
\end{equation}
where \( \alpha \) is the learning rate, and \( \mathcal{L} \) is the task-specific loss function in the new environment. The learning execution and training hyperparameters for RL, TL and CL including exploration strategies and optimization settings, are summarized in Table~\ref{tab:learning_execution_params}.

However, the effectiveness of \ac{TL} largely depends on the quality of the source model. If the chosen source environment is dissimilar to the new one, transfer may lead to poor adaptation or even negative transfer. This brings us to the core of our model selection framework. Model selection is the mechanism that determines which pre-trained model is best suited for transfer to a new environment. This model selection approach has not been previously applied in any \ac{UAV} related \ac{ML} research.To support this, we maintain a library of models \( \{M_1, M_2, \dots, M_n\} \), where each \( M_i \) is trained using \ac{DDQN} in a specific environment \( \mathcal{E}_i \) (e.g., York, London, Ottawa). Each environment is described by a feature vector capturing characteristics such as average building height, building density, station placement, and coverage area. When a \ac{UAV} is deployed in a new environment \( \mathcal{E}_{\text{new}} \), we extract its feature representation and compute a similarity score \( S(\mathcal{E}_{\text{new}}, \mathcal{E}_i) \) for each known environment. Similarity metrics are employed to assess how closely environments align in feature space. However, this computation is challenging, as cities often differ across numerous dimensions that are hard to quantify. To address this, the next section introduces our proposed approach. If the highest similarity score exceeds a predefined threshold \( \tau \), the most similar model is selected for transfer.
\begin{equation}
M^* = 
\begin{cases} 
\arg\max_{i} S(\mathcal{E}_{\text{new}}, \mathcal{E}_i), & \text{if } \max_i S(\mathcal{E}_{\text{new}}, \mathcal{E}_i) \geq \tau, \\
M_G, & \text{otherwise}.
\end{cases}
\end{equation}

Here, \( M_G \) refers to fallback model trained in a synthetic urban environment. The synthetic environment is constructed by aggregating and averaging key structural and deployment parameters extracted from all real city environments available in the model library. 
These parameters include map size, average and maximum building height, building density, \ac{UAV} operating altitude, and terrestrial base station density, as outlined in the system model. Rather than approximating any specific real city, the synthetic environment is deliberately designed to represent a neutral and unbiased urban setting. This design choice prevents the fallback model from overfitting to a particular city morphology and mitigates the risk of negative transfer when the target environment differs significantly from all available source environments.

\begin{figure}[t]
    \centering
    \includegraphics[width=1\columnwidth]{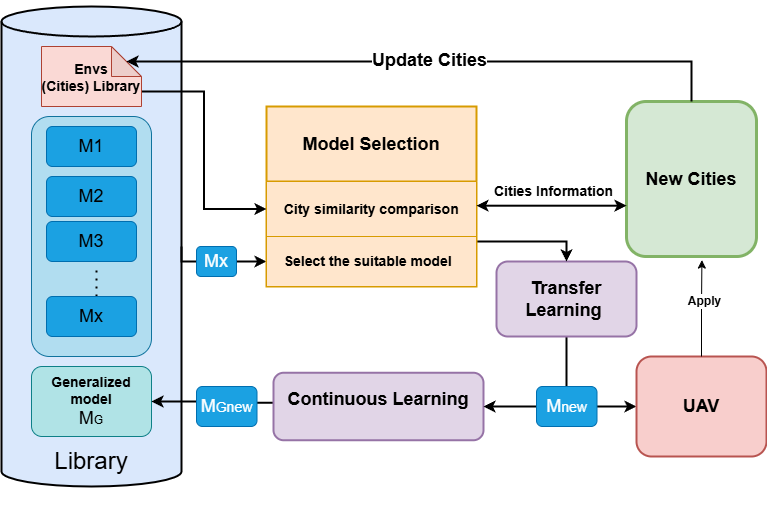}
    \caption{Model selection process includes updating \( M_G \) and library }
    \label{fig:Model selection}
\end{figure}

As a result, \( M_G \) is able to achieve stable baseline performance across a wide range of previously unseen environments, particularly when no sufficiently similar source model is available for transfer. The role of \( M_G \) is not to replace models selected through similarity based transfer learning, but to provide a reliable initialization when the similarity scores between the target environment and all existing library models fall below the predefined threshold. 
Furthermore, the synthetic environment serves as the foundation for continual learning, allowing knowledge acquired from newly trained models in real deployments to be incrementally aggregated into \( M_G \) over time, thereby progressively improving its generalization capability. By incorporating model selection and a synthetic fallback model, the proposed framework avoids blindly transferring knowledge from arbitrarily chosen source environments and instead performs \ac{TL} in a targeted and context-aware manner. This selective reuse of prior knowledge improves adaptation efficiency, reduces the risk of negative transfer, and enhances the scalability and reliability of \ac{UAV} trajectory optimization in heterogeneous urban environments. 
Fig.~\ref{fig:Model selection} illustrates the overall workflow, including model selection, transfer learning, and the update of \( M_G \) and the environment library. 
The following sections detail the proposed city similarity measurement and the continual learning mechanism.

\section{Environment Similarity Comparison for Model selection \ac{TL} } \label{env}

To enable model selection for \ac{UAV} trajectory transfer, we define a similarity function that compares a new environment \( \mathcal{E}_{\text{new}} \) with each pretrained environment \( \mathcal{E}_i \in \mathcal{L} \). The goal is to find the environment most similar to \( \mathcal{E}_{\text{new}} \) based on structural, geographical, and deployment related features, thereby selecting the most suitable model for \ac{TL}. Each environment is described using key similarity factors. These factors capture the dominant environmental factors affecting \ac{UAV} trajectory optimization, including building height and density, terrestrial base station distribution, map scale, and \ac{UAV} operating altitude. Rather than isolating individual factors, the proposed similarity representation aggregates these factors to reflect their combined effect on policy convergence and transfer reliability in realistic urban scenarios. The selection of these factors is guided by prior controlled simulation studies, where individual system variables were independently configured to identify their influence on transfer learning effectiveness \cite{Sun_wincom24}. Building on these insights, this work adopts the selected factors to construct a practical and scalable city similarity mechanism suitable for real world environments.

\subsection{Building Height Similarity}

Let \( H_{\text{new}} \) and \( H_i \) represent the average building heights in \( \mathcal{E}_{\text{new}} \) and \( \mathcal{E}_i \), respectively. Building height is computed by averaging the heights of all 3D building polygons in the environment. The similarity is:
\begin{equation}
\Delta H = |H_{\text{new}} - H_i|,\quad
S_{\text{height}}(\mathcal{E}_{\text{new}}, \mathcal{E}_i) = 1 - \frac{\Delta H}{H_{\max}},
\end{equation}
where \( H_{\max} \) is the maximum observed difference in building height across all environments. Environments with similar urban scale (e.g., both residential or both high-rise) will have a higher similarity score.

\subsection{Building Coverage Similarity}

Building coverage ratio quantifies how much ground space is occupied by buildings. The map is divided into a uniform 2D grid, and the number of occupied cells is counted:
\begin{equation}
C = \frac{N_{\text{covered}}}{N_{\text{total}}},
\end{equation}
where \( N_{\text{covered}} \) is the number of grid cells intersecting with building footprints, and \( N_{\text{total}} \) is the total number of grid cells. The similarity is:
\begin{equation}
\Delta C = |C_{\text{new}} - C_i|,\quad
S_{\text{coverage}}(\mathcal{E}_{\text{new}}, \mathcal{E}_i) = 1 - \Delta C.
\end{equation}
A high similarity score indicates comparable building density and layout patterns.

\subsection{UAV Altitude Similarity}

Each environment uses a fixed \ac{UAV} altitude during flight. Let \( h_{\text{new}} \) and \( h_i \) be the \ac{UAV} operating heights in the new and \( i \)-th environment, respectively:
\begin{equation}
\Delta h = |h_{\text{new}} - h_i|,\quad
S_{\text{uav}}(\mathcal{E}_{\text{new}}, \mathcal{E}_i) = 1 - \frac{\Delta h}{h_{\max}},
\end{equation}
where \( h_{\max} \) is the largest observed difference in \ac{UAV} height across the dataset. Similar \ac{UAV} altitudes often imply comparable LoS probability and \ac{SINR} profiles.

\subsection{Map Size Similarity}

Map size refers to the physical area of the ray-traced region used for \ac{UAV} trajectory simulation. Let \( A_{\text{new}} \) and \( A_i \) be the map areas (in \( \text{m}^2 \)) of the new and \( i \)-th environments, respectively:
\begin{equation}
\Delta A = |A_{\text{new}} - A_i|,\quad
S_{\text{map}}(\mathcal{E}_{\text{new}}, \mathcal{E}_i) = 1 - \frac{\Delta A}{A_{\max}}.
\end{equation}
A similar map size often reflects similar coverage scopes and expected \ac{UAV} travel range.

\subsection{Terrestrial \acp{O-DU} Density Similarity}

Let \( N_{\text{bs}}^{\text{new}} \) and \( N_{\text{bs}}^i \) be the number of terrestrial \acp{O-DU} deployed in the new and \( i \)-th environment. This reflects the deployment density of the network:
\begin{equation}
\Delta N_{\text{bs}} = |N_{\text{bs}}^{\text{new}} - N_{\text{bs}}^i|,\quad
S_{\text{bs}}(\mathcal{E}_{\text{new}}, \mathcal{E}_i) = 1 - \frac{\Delta N_{\text{bs}}}{N_{\text{bs}}^{\max}},
\end{equation}
where \( N_{\text{bs}}^{\max} \) is the maximum number of stations across all environments. A similar terrestrial \acp{O-DU} layout ensures that learned coverage policies transfer more reliably.

\subsection{Composite Similarity Score}

The total similarity score between environments is computed as a weighted sum:
\begin{align}
S(\mathcal{E}_{\text{new}}, \mathcal{E}_i) &= 
w_1 S_{\text{height}} + w_2 S_{\text{coverage}} + w_3 S_{\text{uav}} \nonumber \\
&\quad + w_4 S_{\text{map}} + w_5 S_{\text{bs}},
\end{align}
where \( w_1, \ldots, w_5 \in [0, 1] \) represent the relative importance of each feature. This weighted combination is designed as a  taskoriented similarity abstraction that captures the combined impact of key environmental factors on UAV trajectory learning and transfer effectiveness, rather than as a detailed geometric or topological model of urban environments. The environment \( \mathcal{E}_i \) with the highest similarity score is selected for model transfer.

\section{Update $M_G$ with Continual Learning (CL)}

\begin{algorithm}[ht]
\caption{ Continual Learning for UAV Trajectory Optimization}
\label{algo:FLalgorithm}
\begin{algorithmic}[1]
\State Initialize fallback model $M_G$, new environment model $M_{\text{new}}$, updated model $M_U$, learning rate $\eta$, exploration $\epsilon$, discount factor $\gamma$
\State Distribute $M_G$ to local UAV models $M_{\text{new}}$
\State Initialize replay buffers $\mathcal{R}_G, \mathcal{R}_{\text{new}}$
\For{$t = 1$ to $T$}
    \For{$i \in \{G,\text{new}\}$} \Comment{Train each local model}
        \State Reset environment $i$, initialize state $s_i$
        \For{episode = 1 to $N_{\text{epi}}$}
            \For{$t = 0$ to $N_{\text{step}}$}
                
                \State Execute $a_t$, observe next state $s_{t+1}$, reward $r_t$
                \State Store transition $(s_t, a_t, r_t, s_{t+1})$ in $\mathcal{R}_i$
                \State Sample minibatch from $\mathcal{R}_i$, update $M_i$
                \State Update target network if necessary
                \State $s_t \leftarrow s_{t+1}$
            \EndFor
        \EndFor
    \EndFor

    \If{$t \mod E == 0$} \Comment{Federated Model Update}
        \State UAVs send local model parameters to the global server
        \State Compute weighted aggregation:
        \[
        M_U \leftarrow w_G M_G + w_{\text{new}} M_{\text{new}}
        \]
        \State Set updated model as new : $M_G \leftarrow M_U$
        \State Synchronize models: $M_{\text{new}} \leftarrow M_G$
    \EndIf

    \If{$t \mod 100 == 0$} \Comment{Performance Evaluation}
        \State Test $M_G$ on both environments, log results
    \EndIf

    \State Decay exploration rate: $\epsilon \leftarrow \max(\epsilon \cdot \text{decay factor}, \epsilon_{\text{min}})$
\EndFor

\State \textbf{return} Final updated $M_G$
\end{algorithmic}
\end{algorithm}

To enhance the effectiveness of the backup model \( M_G \), we employ \ac{CL} to update it which is described in Algorithm~\ref{algo:FLalgorithm}. The model \( M_G \) is initially trained in a simulated environment. When a \ac{UAV} encounters a new environment, a new model \( M_{\text{new}} \) is created through \ac{TL}. While \( M_{\text{new}} \) is stored in the model library for future selection, it is also used to refine \( M_G \). This allows \( M_G \) to continuously accumulate knowledge from diverse environments, improving its generalization over time.

During training, the \ac{UAV} starts from an initial state \( s_t \) and selects an action \( a_t \) using an \(\epsilon\)-greedy policy. The action is executed, leading to a new state \( s_{t+1} \) and a corresponding reward \( r_t \). This transition is stored in a replay buffer and used to update \( M_{\text{new}} \). The training process iterates over multiple episodes, gradually improving the model's ability to optimize \ac{UAV} trajectories. After local training, the fallback model is updated using a weighted aggregation strategy. Instead of equal-weight averaging, the weights assigned to \( M_G \) and \( M_{\text{new}} \) are adjusted based on the similarity between the new environment and the original training conditions. The updated model is computed as
\begin{equation}
M_G' \leftarrow w_0 M_G + w_{\text{new}} M_{\text{new}},
\end{equation}
where the weights are determined by an environment similarity metric \( S_{\text{env}} \),
\begin{equation}
w_0 = \frac{S_{\text{env}}}{S_{\text{env}} + 1}, \quad w_{\text{new}} = \frac{1}{S_{\text{env}} + 1}.
\end{equation}
If the new environment closely resembles the original, the weight of \( M_G \) remains dominant, ensuring knowledge retention. Conversely, if the environment is significantly different, the contribution of \( M_{\text{new}} \) is increased, allowing greater adaptation to new conditions. The updated \( M_G' \) is then redistributed to \acp{UAV} for further training, ensuring consistency across deployments. The exploration rate \( \epsilon \) is gradually reduced over time to encourage exploitation of learned policies. Periodic evaluations are conducted on both the simulation and real-world environments to assess performance and verify improvements in trajectory optimization.

The proposed approach allows \acp{UAV} to dynamically balance knowledge retention and adaptation, preventing catastrophic forgetting while enabling continual learning. By leveraging \ac{CL} with adaptive weighting, \ac{UAV}-based \ac{O-RAN} systems can improve trajectory planning in diverse and evolving conditions while maintaining efficiency and robustness. To evaluate this framework, we conduct simulations across multiple real-world environments.

\begin{figure*}[t]
    \centering
    \includegraphics[width=1\textwidth]{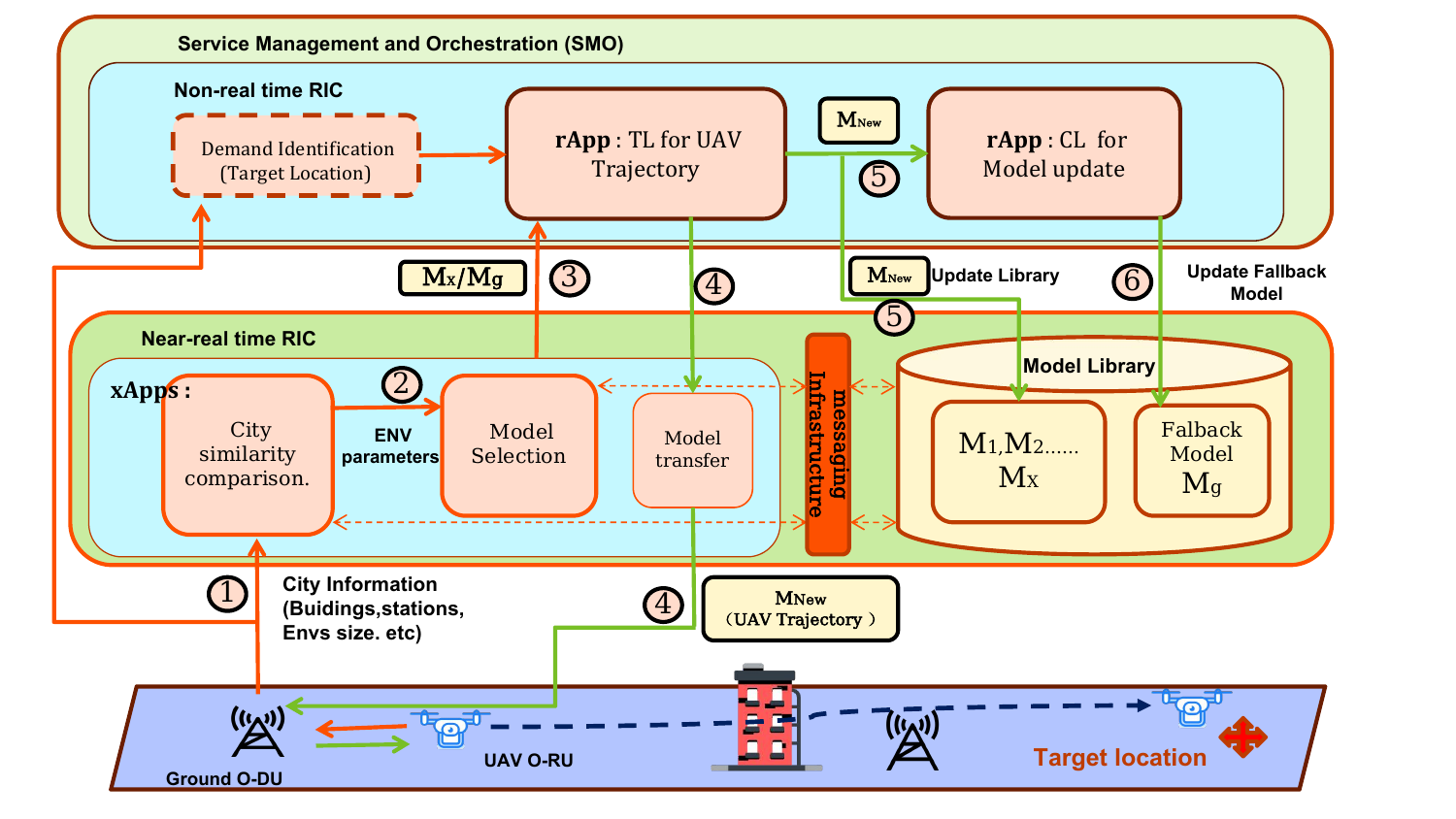}
    \caption{Workflow of UAV trajectory optimization in the O-RAN framework.}
    \label{fig:workflow}
\end{figure*}

\begin{table}[t]
\centering
\small
\caption{Summary of System and Ray-Tracing Parameters}
\label{tab:parameters}
\begin{tabular}{lll}
\hline
\textbf{Parameter Name} & \textbf{Letter} & \textbf{Value} \\
\hline

Launch Area                       & \( \mathcal{A}_{\text{launch}} \)  & 200m × 200m \\
Target Area                       & \( \mathcal{A}_{\text{target}} \)  & 400m × 400m \\
Grid Resolution                   & \( r \)                            & 5m \\
BS Transmit Power                 & \( P_t \)                          & 30 dBm \\
SINR outage Threshold             & \( \phi_{\text{th}} \)             &  0 dB \\
Max Steps                         & \( K_{\max} \)                     & 200 \\
Max UAV Velocity                  & \( V_{\max} \)                     & 20 m/s  \\
Altitude Bounds                   & \( [h_{\min}, h_{\max}] \)         & [40m, 200m] \\
Weighting Factor                 & \( w_r \)                            & 0.8 \\
Reflections (Ray Tracing)        & \( R \)                            & 3 \\
Diffractions (Ray Tracing)       & \( D \)                            & 1 \\
Carrier Frequency                & \( f_c \)                          & 450 MHz \\
System Bandwidth                 & \( B \)                            & 1 MHz \\
\hline
\end{tabular}
\end{table}

\begin{table}[ht]
\centering
\caption{Learning and Training Parameters for RL, TL, and CL}
\label{tab:learning_execution_params}
\begin{tabular}{l c}
\hline
\textbf{Parameter} & \textbf{Value} \\
\hline
Learning rate ($\alpha$) & 0.001 \\
Initial exploration rate (DDQN) ($\epsilon$) & 1.0 \\
Exploration decay rate ($\epsilon_{\text{decay}}$) & 0.998 \\
Initial exploration rate (Transfer) ($\epsilon_{\text{transfer}}$) & 0.5 \\
Step penalty ($R_n$) & 1 \\
Reward for reaching target ($\mathcal{R}_{\text{goal}}$) & 2000 \\
Outage penalty ($\mathcal{C}_{\text{out}}$) & 10000 \\
\hline
Maximum training episodes (RL) & 5000 \\
Maximum training episodes (TL) & 2000 \\
Replay buffer size (RL/TL) & 100000 \\
Minimum replay size (RL/TL) & 5000 \\
Mini-batch size & 32 \\
Target network update interval & 5 episodes \\
Multi-step return length (RL/TL) & 30 \\
Maximum steps per episode & 100 \\
Action space size & 4 \\
\hline
Federated learning rounds (CL) & 1000 \\
Local episodes per round & 10 \\
Total local episodes & 10000 \\
Number of clients & 2 (Beijing, York) \\
Aggregation method & FedAvg \\
Minimum replay size (CL) & 1000 \\
Multi-step return length (CL) & 10 \\
\hline
\end{tabular}
\end{table}

\section{O-RAN-Based UAV Adaptive System}\label{Oran}

We consider a \ac{UAV} trajectory optimization framework that integrates \acp{UAV} as \ac{O-RAN} \acp{RU}, maintaining connectivity with ground \ac{O-DU}. The system is built upon the \ac{O-RAN} architecture where both the \ac{Non-RT RIC} and \ac{Near-RT RIC} are hosted within the ground-based DU/CU infrastructure. These RICs facilitate the storage, training, and deployment of \ac{UAV} trajectory models. Through enhanced \ac{TL}, the system enables efficient model adaptation in new environments, while \ac{CL} ensures continued refinement of the fallback models. As illustrated in Figure~\ref{fig:workflow}, this architecture effectively combines \ac{TL} and \ac{CL} to improve the scalability, responsiveness, and planning accuracy of \acp{UAV} across diverse

The service management and orchestration (SMO) framework hosts the Non-RT RIC, which handles long-term policy management and \ac{ML} tasks. Within the Non-RT RIC, dedicated rApps are implemented for \ac{TL} and \ac{CL}. The \ac{TL} rApp is responsible for training \ac{UAV} trajectory models to adapt to new environmentsand stores the resulting model in the shared library. Subsequently, the \ac{CL} rApp aggregates these newly added models to update the fallback model  \( M_G \) by aggregating knowledge from newly trained models. The Near-RT RIC, operating on a timescale of 10 ms to 1 s, manages time-sensitive functions such as real-time trajectory selection, demand identification, and model dispatching. It contains a model library that stores all pre-trained models \( M_1, M_2, ..., M_X \), as well as the fallback model \( M_G \). Two dedicated xApps operate within the Near-RT RIC: a city similarity comparison xApp and a model selection xApp. The former computes structural similarity between the new environment and known cities, while the latter selects the most appropriate model from the library based on similarity scores.

The \ac{UAV} trajectory optimization process begins by identifying new environmental information. When a \ac{UAV} is deployed in a new environment, the city similarity comparison xApp compares the extracted map features with those in the database. If a sufficiently similar environment is found, the corresponding pre-trained model is selected for \ac{TL} and apply to the \ac{UAV}. Otherwise, the \( M_G \) is used. City similarity comparison, model selection, and model library exchanges occur dynamically in real time within the Near-RT RIC. Once the selection is made, the \ac{TL} rApp in the Non-RT RIC initiates fine-tuning of the selected model (or \( M_G \)) based on the newly gathered environmental information and destination targets. The new trained model from \ac{TL} will be uploaded to the model library and transferred to the \ac{UAV} through Near-RT RIC. Simultaneously, newly trained models and associated map data are added to expand the model library, ensuring growing coverage and accuracy. To maintain and enhance the versatility of the \( M_G \) over time, each newly trained trajectory model is used to incrementally update \( M_G \) via \ac{CL} implemented in the \ac{CL} rApp. This guarantees that \( M_G \) continuously evolves by incorporating knowledge from diverse environments while preserving broad generalization capabilities.

\section{Simulation Results and Analysis }
\begin{table*}[ht]
\centering
\caption{City Map Information}
\label{tab:city_map_info}
\begin{tabular}{|l|r|r|r|r|r|r|}
\hline
\textbf{City}       & \textbf{Area (m²)} & \textbf{Building Coverage (\%)} & \textbf{Avg Height (m)} & \textbf{Max Height (m)} & \textbf{Station Count} & \textbf{UAV Height (m)} \\ \hline
Birmingham  & 769034.69   & 59.25  & 22.61  & 91.17  & 4  & 100  \\ \hline
Manchester  & 1032888.38  & 58.83  & 15.81  & 96.44  & 5  & 100  \\ \hline
York        & 629907.38   & 49.18  & 9.82   & 22.42  & 8  & 50   \\ \hline
Beijing     & 558407.31   & 47.15  & 10.38  & 23.09  & 3  & 40   \\ \hline
London      & 1290853.62  & 70.94  & 23.77  & 292.94 & 10 & 200  \\ \hline
Rosslyn     & 305200.00   & 68.70  & 31.20  & 123.40 & 3  & 140  \\ \hline
Ottawa      & 604800.00   & 71.70  & 15.30  & 52.00  & 4  & 70   \\ \hline
\end{tabular}
\end{table*}

\begin{table*}[ht!]
\centering
\caption{Similarity Scores of maps in Library to Target maps}
\begin{tabular}{|l|c|c|c|c|c|c|c|}
\hline
\textbf{Reference City}&\textbf{Birmingham} & \textbf{Manchester} & \textbf{London} & \textbf{Rosslyn} & \textbf{York} & \textbf{Ottawa} &\textbf{Beijing}\\ \hline
\textbf{York (1st New Env)}           & \textbf{0.548}               & 0.536               & 0.456         & 0.44             & 1.000        & -        & -    \\ \hline
\textbf{Ottawa (2nd New Env)}         & 0.682               & \textbf{0.729}               & 0.642         & 0.610           & 0.648         & 1.000      & -    \\ \hline
\textbf{Beijing (3rd New Env)}        & 0.641               & 0.628               & 0.461         & 0.459           & \textbf{0.837}         & 0.703    & 1.000   \\ \hline
\end{tabular}
\label{tab:similarity}
\end{table*}

\subsection{Environment Setup}

To evaluate the proposed \ac{UAV} trajectory optimization framework, we employ Wireless InSite, a high-fidelity wireless communication simulation software developed by Remcom \cite{remcom2023wireless}. Wireless InSite uses ray-tracing and empirical models to provide accurate radio propagation predictions, making it suitable for modeling realistic urban environments. To ensure diverse and realistic evaluation of \ac{UAV}-based \ac{O-RAN} deployments, we conduct simulations across multiple urban scenarios, including York, Manchester, London, Birmingham, Rosslyn, Ottawa, and Beijing, as illustrated in Fig.~\ref{fig:maps}. Each selected map represents a unique urban landscape with distinct architectural and geographical characteristics. The London map focuses on the city center, where many buildings exceed typical \ac{UAV} flight altitudes, requiring careful navigation around tall structures. In contrast, Manchester and Birmingham are modeled in urban but non-central areas that feature mixed environments, including mid-rise buildings and residential zones. York presents a low-rise cityscape due to historical preservation policies that limit building height. An especially unique case is the map of Beijing, which captures the Forbidden City, an ancient imperial complex characterized by its expansive layout and uniformly low building profiles, despite being located at the heart of a modern metropolis.

Wireless InSite is used to generate high-fidelity radio maps for each city, providing realistic channel modeling that incorporates diffraction, reflection, and scattering effects. The tool supports multi-frequency propagation analysis across sub-6 GHz and mmWave bands and enables dynamic \ac{UAV} simulation to accurately capture mobility-induced variations in wireless signal behavior. These features ensure that each simulated environment contributes meaningfully to assessing the adaptability and robustness of \ac{UAV} trajectory learning systems under different urban conditions. To generate urban maps for Wireless InSite, we use two methods: the first is \ac{OSM} + QGIS, which extracts city data, converts it into a \texttt{.city} file, and allows material property adjustments. Another method is using 3DCITYLoader, which directly downloads 3D city models in \texttt{STL} format, which can be imported along with terrain data. By integrating these high-fidelity simulations, we ensure an accurate evaluation of \ac{UAV} trajectory planning and wireless performance in realistic urban settings.

Table~\ref{tab:city_map_info} provides a comparative analysis of key urban characteristics across multiple cities, including total area, building coverage percentage, average and maximum building heights, station count, and suggested \ac{UAV} operational altitude. The area represents the total land covered by the city model, while the building coverage percentage indicates the proportion of the city occupied by structures. Average and maximum building heights provide insight into the vertical profile of each city, which is essential for \ac{UAV} flight planning and communication infrastructure deployment. The \ac{UAV} height, calculated as the maximum building height plus an additional buffer, ensures safe and effective aerial operations. 

\begin{figure*}[ht]
\centering

\begin{subfigure}{0.3\textwidth}
    \centering
    \includegraphics[width=\linewidth]{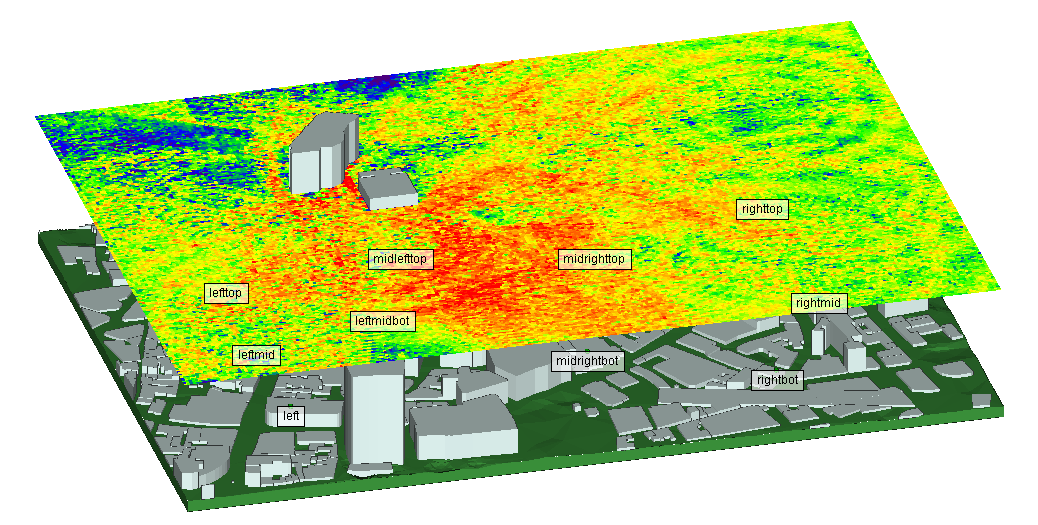}
    \caption{London}
\end{subfigure}
\hfill
\begin{subfigure}{0.3\textwidth}
    \centering
    \includegraphics[width=\linewidth]{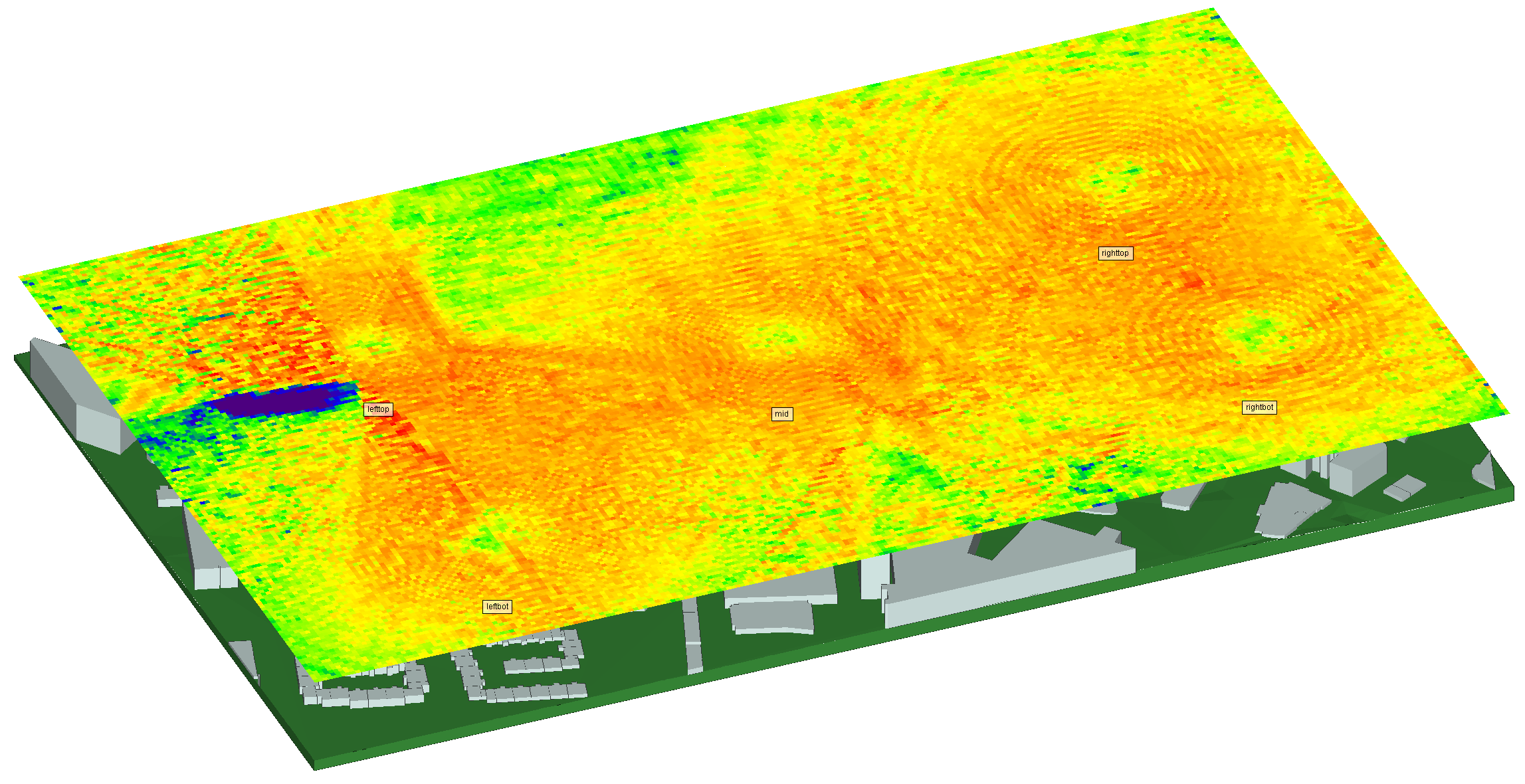}
    \caption{Manchester}
\end{subfigure}
\hfill
\begin{subfigure}{0.3\textwidth}
    \centering
    \includegraphics[width=\linewidth]{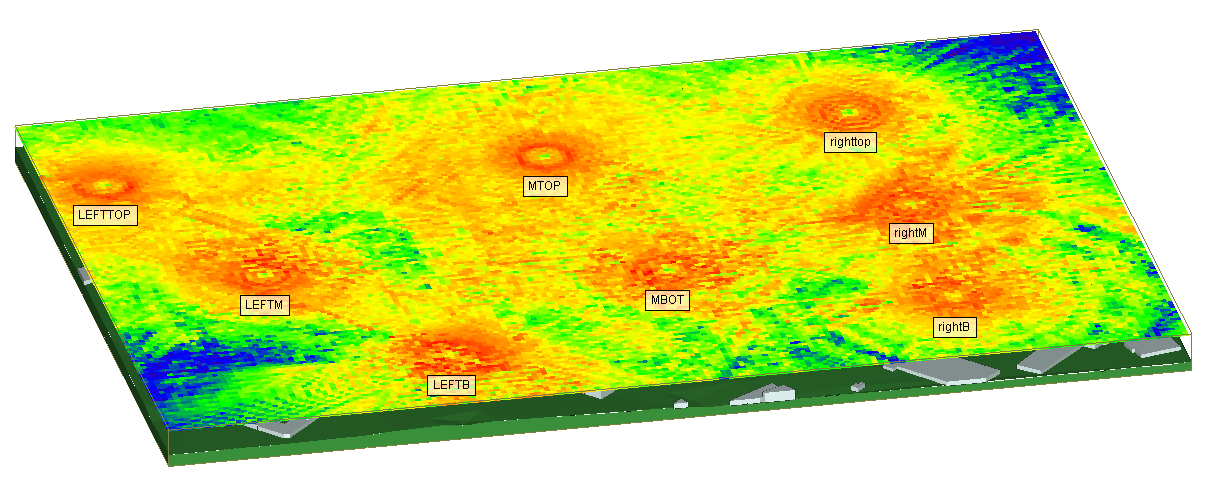}
    \caption{York}
\end{subfigure}
\begin{subfigure}{0.3\textwidth}
    \centering
    \includegraphics[width=\linewidth]{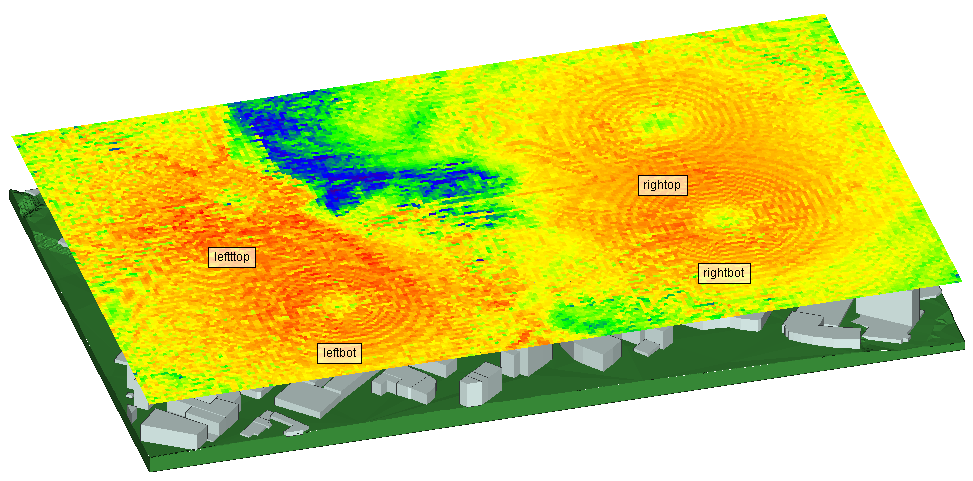}
    \caption{Birmingham}
\end{subfigure}
\hfill
\begin{subfigure}{0.3\textwidth}
    \centering
    \includegraphics[width=\linewidth]{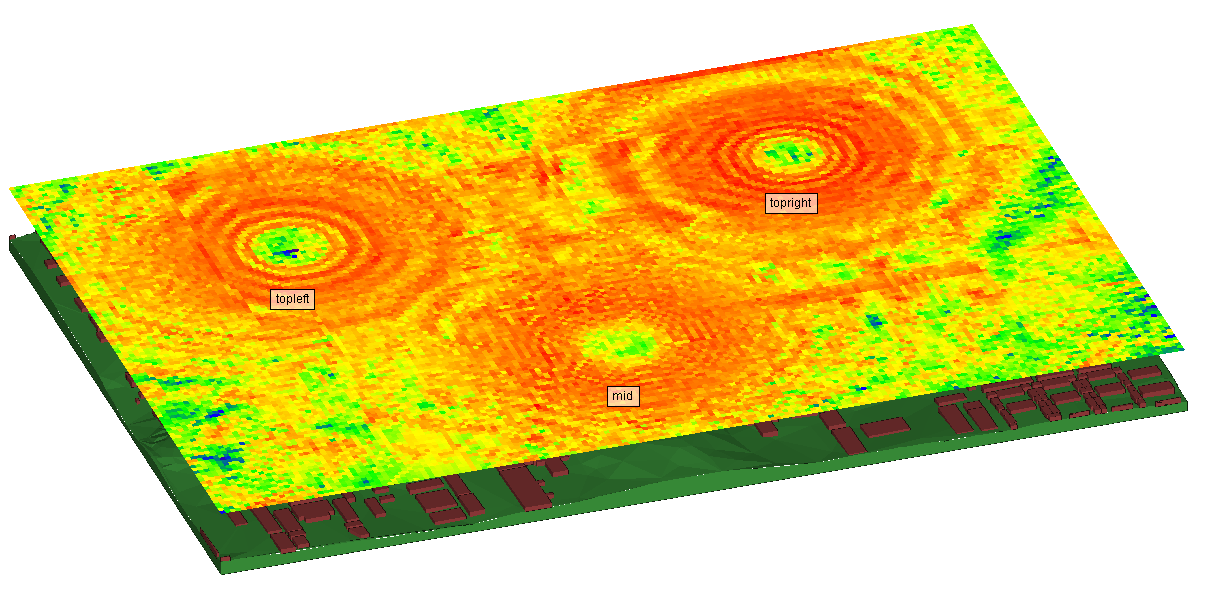}
    \caption{Beijing}
\end{subfigure}
\hfill
\begin{subfigure}{0.3\textwidth}
    \centering
    \includegraphics[width=\linewidth]{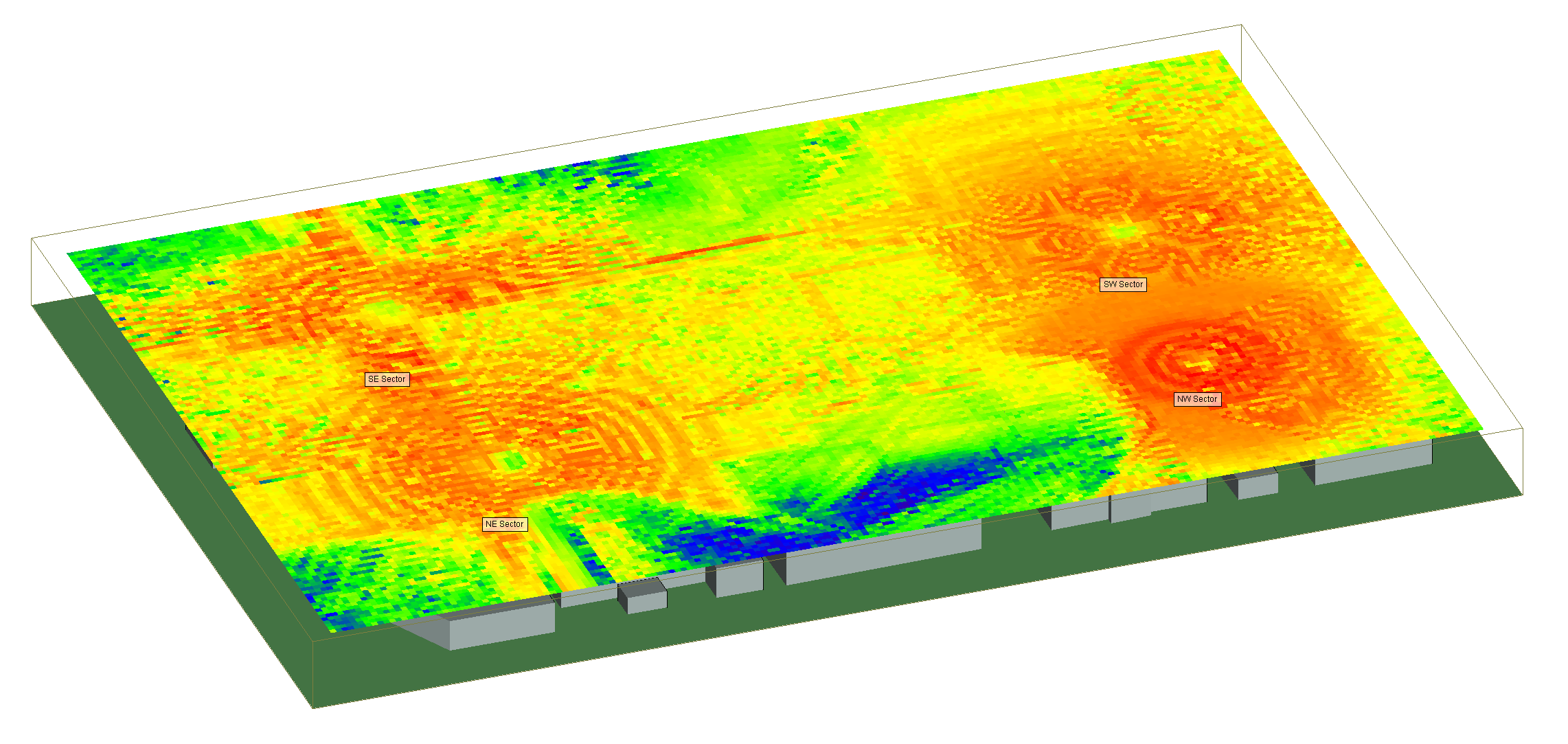}
    \caption{Ottawa}
\end{subfigure}
\caption{SINR radio maps generated by ray tracing in different cities, showing the \ac{UAV} flight altitude. Cold colors such as blue indicate poor signal and high probability of disconnection. Warm colors such as red indicate good signal and low probability of disconnection, which are ideal areas for \ac{UAV} flight.}
\label{fig:sixradiomaps}
\end{figure*}

\subsection{Generation of City Radio Map}

To enable trajectory learning and evaluation, it is essential to provide the \ac{UAV} with realistic spatial information about signal availability and outage likelihood across the operating environment. The radio map serves as a critical input for training, offering localized communication quality data that allows the \ac{UAV} to make informed decisions during path planning and avoid regions with high outage probability.

The simulation map is generated using Wireless InSite and includes detailed material properties to enhance propagation accuracy. These materials include dielectric half-space types (e.g., concrete, wet earth) and one-layer dielectric structures (e.g., brick), enabling the simulation to more accurately capture real-world signal behavior. The terrestrial \acp{O-DU} is configured with four half-wave dipole antennas and a transmit power of 30 dBm, while the \ac{UAV} operates at a fixed altitude of 80 meters to ensure consistent evaluation. To model the ground-to-air communication characteristics at \ac{UAV} height, we deploy a dense grid of receivers at 80 meters altitude with 5-meter spacing between adjacent points. This setup creates a structured dataset of \(192 \times 126\) points, effectively capturing \ac{SINR} variations throughout the flight zone. However, because \acp{UAV} may fly through regions not directly sampled in the grid, gaps in the data can occur. To address this, we apply an interpolation technique that fills missing values using the median of neighboring points, ensuring spatial continuity.

The processed data is then rescaled to the physical dimensions of the simulation area (960 meters by 630 meters). Finally, \ac{SINR} values are converted into outage rate values, providing a more interpretable and policy-relevant measure of connectivity quality. As shown in Fig.~\ref{fig:sixradiomaps}, the resulting radio map visually depicts communication quality, where red areas indicate high \ac{SINR} and low outage rates, and blue areas correspond to low \ac{SINR} and high outage zones. This radio map becomes the foundation for environment modeling and policy learning in all subsequent trajectory simulations.

\section{Results}

We now present the results of the city comparison, model selection process, and the impact of \ac{CL} in updating the \( M_G \). The study begins with York as the first new environment, where four existing environments and their trained models are used for \ac{TL}. After evaluating model selection in York, it is added to the library, expanding the knowledge base to five environments before transitioning to Ottawa. The process is then repeated, incorporating Ottawa into the library, leading to a final evaluation in Beijing with six environments available for transfer. The results examine whether the model selection framework makes optimal choices for \ac{TL} and how much it improves performance compared to \ac{DDQN} and conventional \ac{TL} methods. The key metrics analyzed include training speed and energy cost reduction. Additionally, the \ac{CL} framework is evaluated for updating the fallback model \( M_G \), demonstrating how \ac{CL} enhances its adaptability in dynamic urban environments. Table~\ref{tab:outage_results} summarizes the lowest stable average outage rates and convergence episodes achieved by \ac{DDQN} and the best-performing transfer models across three different environments. The results highlight the advantage of model selection, where transferring from structurally similar environments significantly reduces convergence time while maintaining low outage rates.

\subsection{Outage Rate Performance Results}
Table~\ref{tab:outage_results} summarizes the outage rate performance of the best models across the new environments. For each target environment, we report the lowest average outage rate achieved, the convergence episodes required by the \ac{DDQN} baseline, and the best-performing \ac{TL} model along with its convergence episodes. In the York environment, the target outage rate of 0.35 was achieved by the $M_G$ in just 880 episodes, compared to the \ac{DDQN} baseline's 1380 episodes. Similarly, for Ottawa, the Manchester transfer model significantly outperformed the baseline, reaching the target outage rate of 0.31 in 750 episodes versus 1340 for \ac{DDQN}. In Beijing, \ac{TL} from York enabled the \ac{UAV} to achieve the target outage rate of 0.38 in only 590 episodes, whereas \ac{DDQN} required 1090 episodes. These results directly address the challenge of optimizing \ac{UAV} trajectories in unfamiliar environments. By leveraging \ac{TL}, the \acp{UAV} rapidly adapt to new scenarios with minimal environmental data, achieving both low outage rates and faster convergence. This demonstrates that \ac{TL} not only accelerates learning but also enhances performance, effectively solving the problem of dynamic trajectory optimization under limited prior knowledge.

\subsection{Model Selection}
\begin{table*}[ht!]
\centering
\caption{Outage Rate Performance of Best Model for Each New Environment}
\label{tab:outage_results}
\begin{tabular}{|l|c|c|c|c|}
\hline
\textbf{Environment} & \textbf{Best Transfer Model} & \textbf{Target (Lowest) Avg Outage Rate} & \textbf{DDQN Convergence Episodes} & \textbf{Best Transfer Convergence Episodes} \\
\hline
York    & \(M_G\)       & 0.35 & 1380 & 880 \\ 
\hline
Ottawa  & Manchester    & 0.31 & 1340 & 750 \\
\hline
Beijing & York          & 0.38 & 1090  & 590 \\
\hline
\end{tabular}
\end{table*}

\subsubsection{Model Selection and Transfer Learning to York}

\begin{figure}[h!]
    \centering
    \includegraphics[width=1\columnwidth]{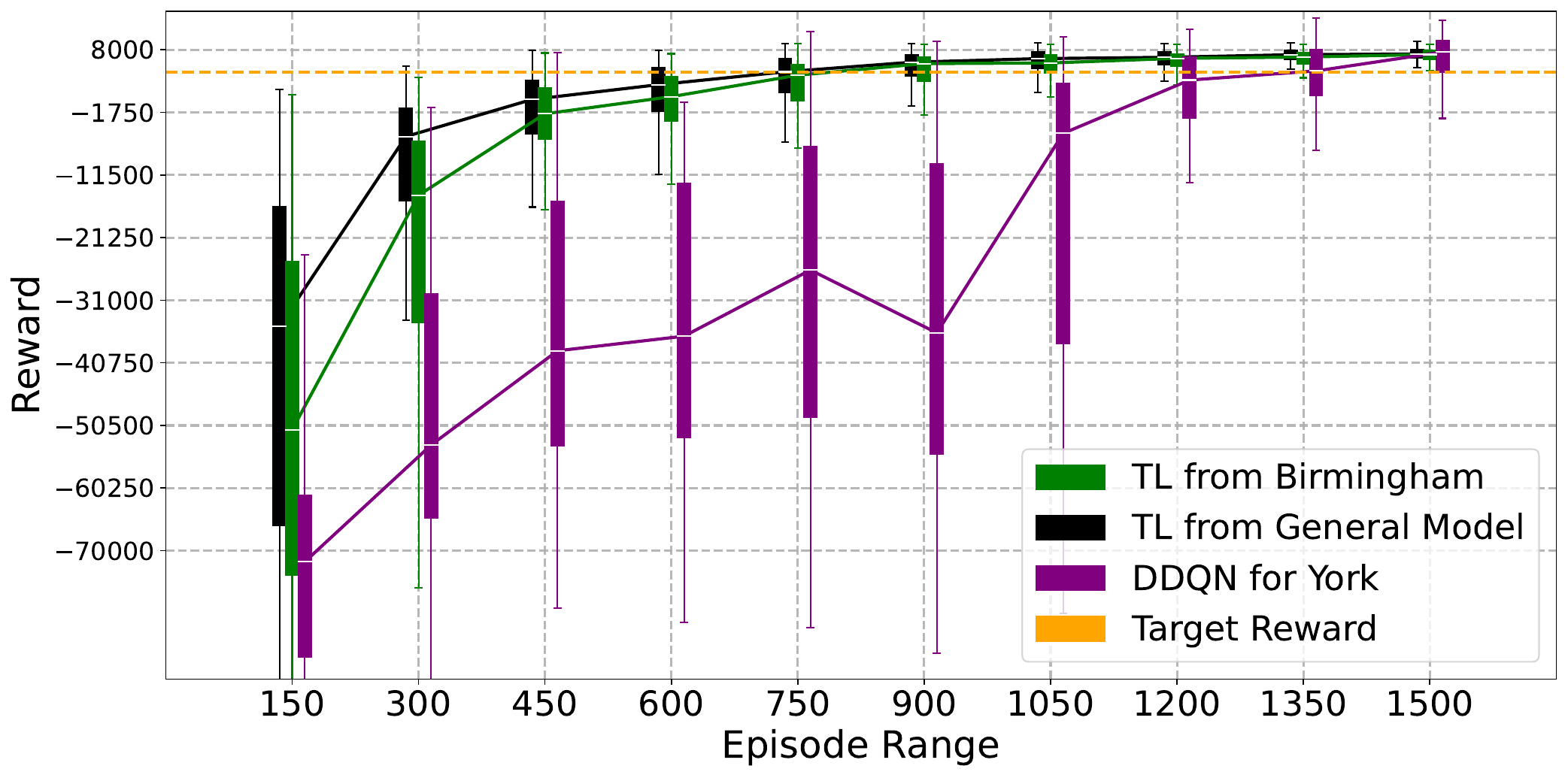}
    \caption{Performance comparison of Model selection, Baseline model,and DDQN models for York.}
    \label{fig:transfer_york}
\end{figure}

\begin{figure}[h]
\centering
\begin{subfigure}[t]{0.8\columnwidth}
    \centering
    \includegraphics[width=\linewidth]{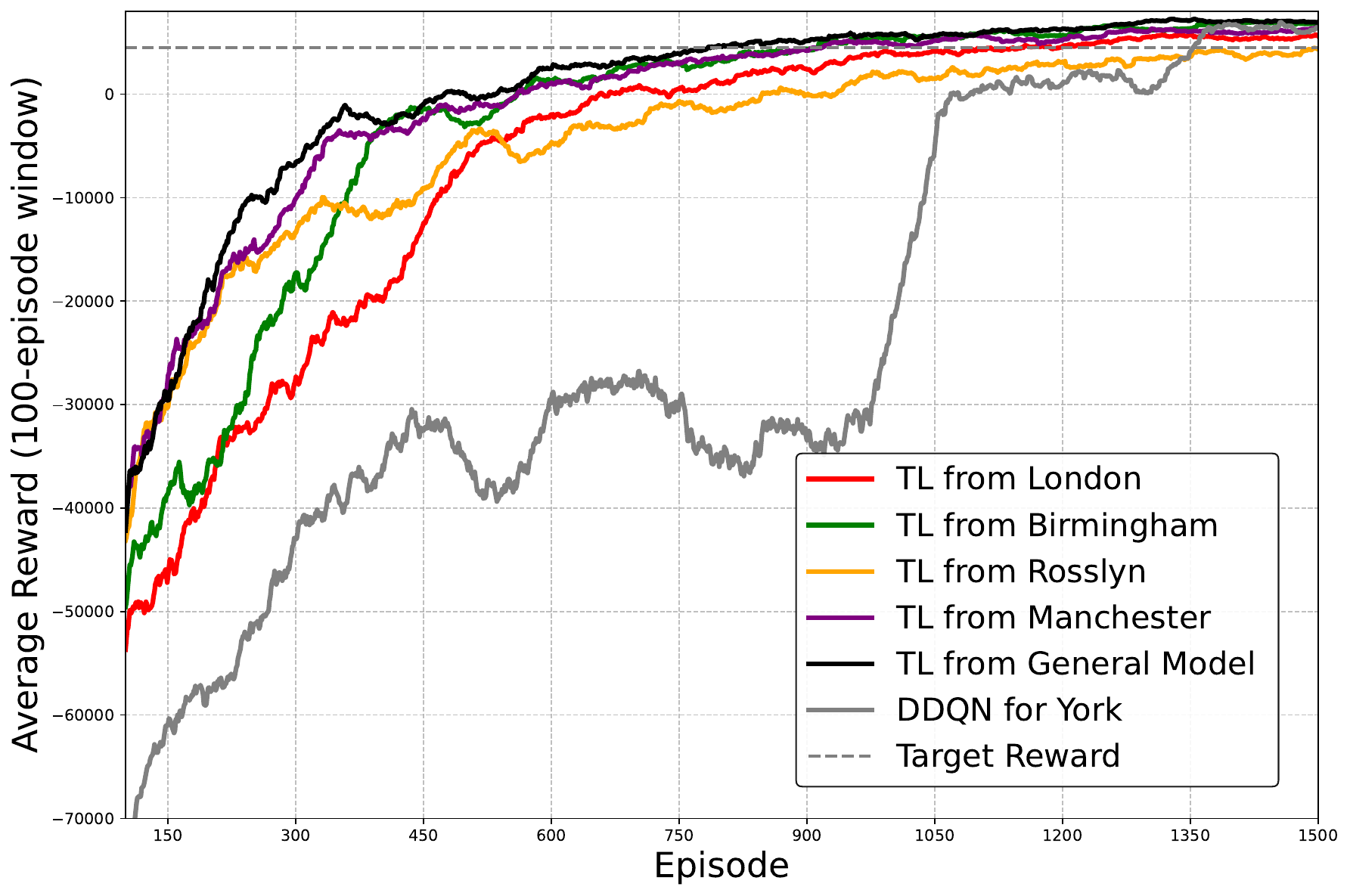}
    \caption{Full training reward curves for all compared models in the York environment.}
    \label{fig:york_full_reward}
\end{subfigure}
\vspace{0.5em}
\begin{subfigure}[t]{0.8\columnwidth}
    \centering
    \includegraphics[width=\linewidth]{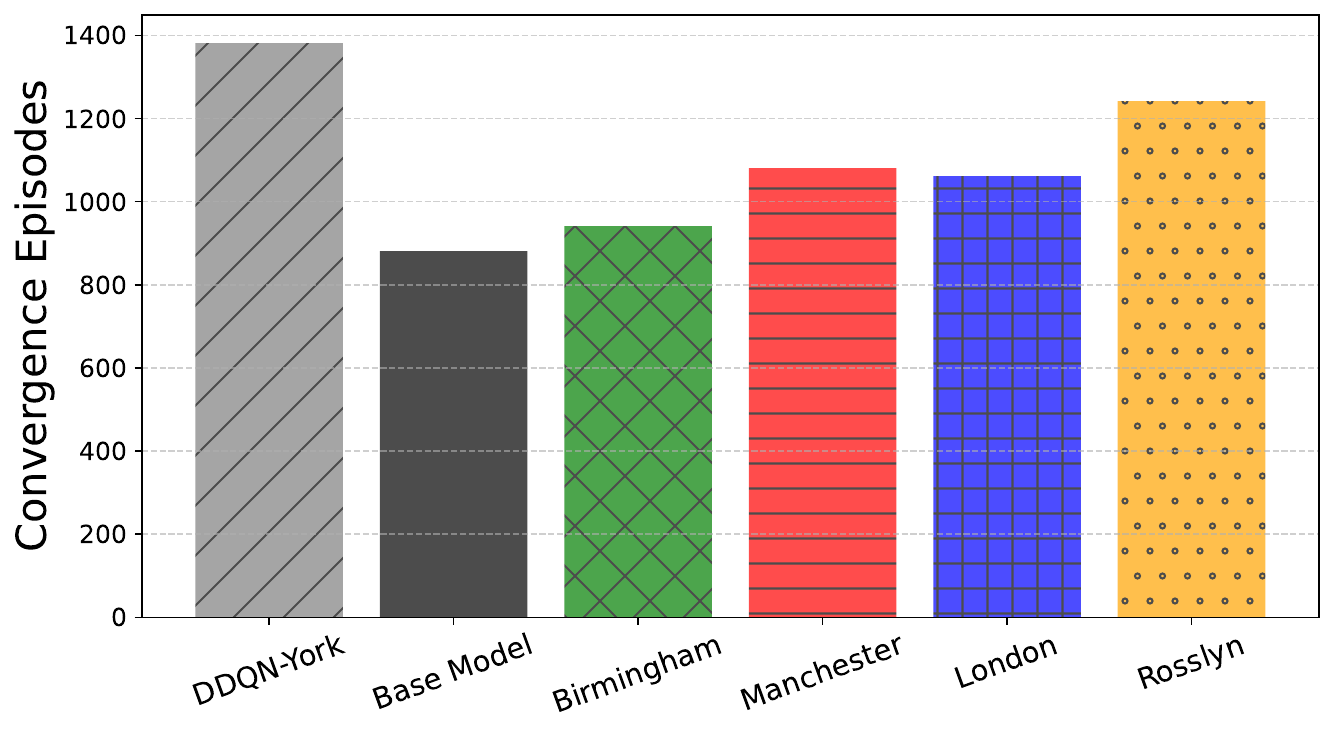}
    \caption{Performance comparison of different source models from the library in York.}
    \label{fig:transferyork}
\end{subfigure}
\caption{York environment: complete training reward curves and convergence performance comparison }
\label{fig:compareyork}
\end{figure}

For the task of transferring to the York environment, model performance was evaluated using both direct transfer from the library and the base \(M_G\), with results shown in Figures~\ref{fig:compareyork} and~\ref{fig:transfer_york}. The similarity scores for York, presented in Table~\ref{tab:similarity}, reveal that Birmingham (0.548) and Manchester (0.536) had the highest similarity among available environments. However, all similarity scores remained below the commonly accepted threshold (0.6), suggesting limited potential for effective direct transfer.

Figure~\ref{fig:compareyork} shows the convergence episode count for each library model, where lower values indicate faster adaptation. The fallback model \(M_G\) achieved convergence at approximately 880 episodes, outperforming models from Birmingham (940), Manchester (1100), London (1070), and Rosslyn (1250). Compared to the best library model which transfer from Birmingham, \(M_G\) reduced convergence time by roughly 6.4\%, and when compared to the worst case which is model from Rosslyn, by over 29.6\%. For reference, our experiments were conducted on a system equipped with an Intel Core i9-12900H CPU and an NVIDIA GeForce RTX 3070 Ti Laptop GPU. Based on this setup, the observed reduction corresponds to a training time savings of approximately 2 hours. This rough estimate is provided to give practical context for understanding the efficiency gains and it is not necessarily a very accurate time or energy improvement measurement. Figure~\ref{fig:transfer_york} includes the performance of the retrained \ac{DDQN} model in York, which required around 1380 episodes to converge. This highlights that the \(M_G\), despite not being trained specifically for York, outperformed 36.2\% faster than training from scratch. The consistent poor performance of models across all libraries reinforces the limitations of traditional \ac{TL} when the target environment differs significantly from the available sources. In contrast, the  \(M_G\) served effectively as a fallback, offering both improved convergence and stability. These findings underscore the importance of complementing model selection strategies with a robust and continually updated fallback model to ensure reliable adaptation in edge cases. Maintaining such a model ensures that \acp{UAV} can still be effectively deployed in unseen environments, even in the absence of high-similarity city data.

\subsubsection{Model Selection and Transfer Learning to Ottawa}

\begin{figure}[h]
\centering
\begin{subfigure}[t]{0.8\columnwidth}
    \centering
    \includegraphics[width=\linewidth]{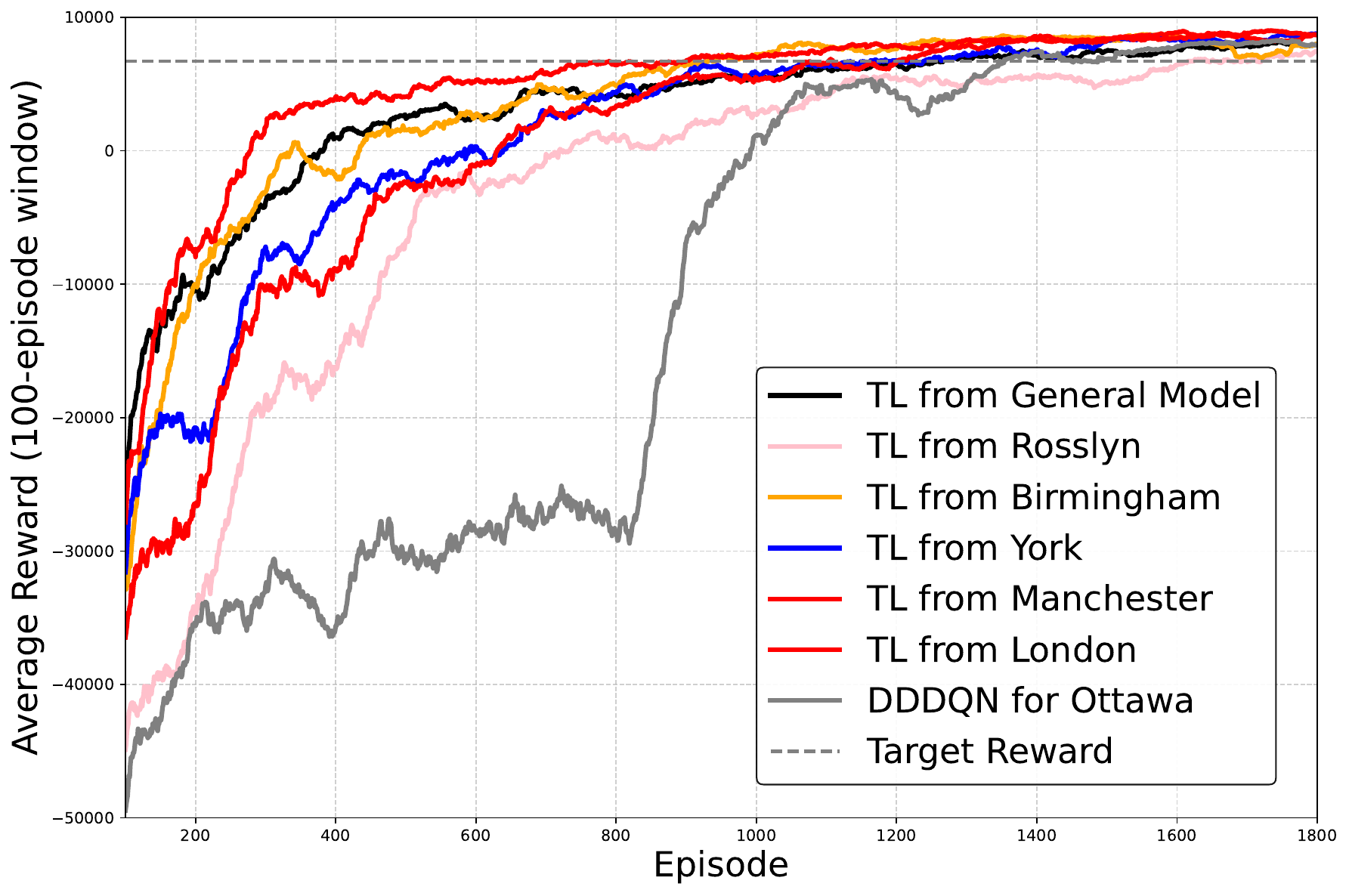}
    \caption{Full training reward for all compared models in the Ottawa environment.}
    \label{fig:ottawa_full_reward}
\end{subfigure}
\vspace{0.5em}
\begin{subfigure}[t]{0.8\columnwidth}
    \centering
    \includegraphics[width=\linewidth]{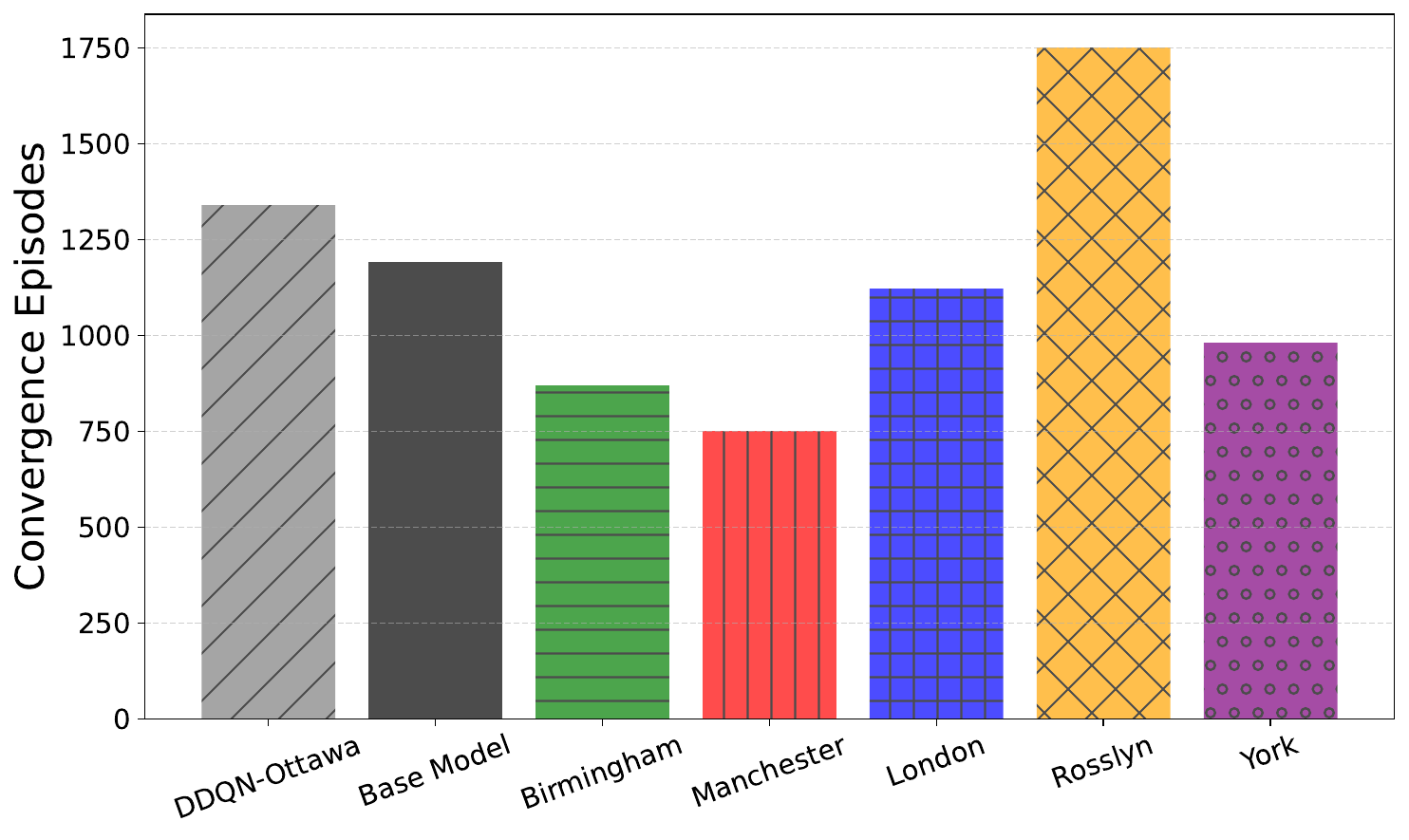}
    \caption{Performance comparison of different source models from the library in Ottawa.}
    \label{fig:compareottawabar}
\end{subfigure}
\caption{Ottawa environment: complete training reward curves and convergence performance comparison across different source models.}
\label{fig:compareottawa}
\end{figure}

To evaluate the transferability of pre-trained models to the Ottawa environment, Figure~\ref{fig:ottawa_full_reward} presents the convergence performance of models selected from the environment library, now expanded to include York. The figure compares convergence episodes across transferred models, the base \(M_G\), and a freshly trained \ac{DDQN} in Ottawa.

Table~\ref{tab:similarity} shows that Manchester has the highest similarity score to Ottawa (0.729), followed by Birmingham (0.682) and York (0.837). Despite this, Figure~\ref{fig:compareottawa} reveals notable differences in convergence behavior. The model from Manchester converged the fastest at around 750 episodes, followed by Birmingham (880) and York (980). In contrast, the base model \(M_G\) required approximately 1180 episodes to converge, and the \ac{DDQN} trained from scratch in Ottawa took around 1340 episodes. The model transferred from Rosslyn was the slowest to converge (1650 episodes), emphasizing its limited suitability for transfer. This result shows that transferring a model from an environment that does not share enough similarity can degrade performance which performing even worse than training from scratch with \ac{DDQN}.

\begin{figure}[ht!]
    \centering
    \includegraphics[width=1\columnwidth]{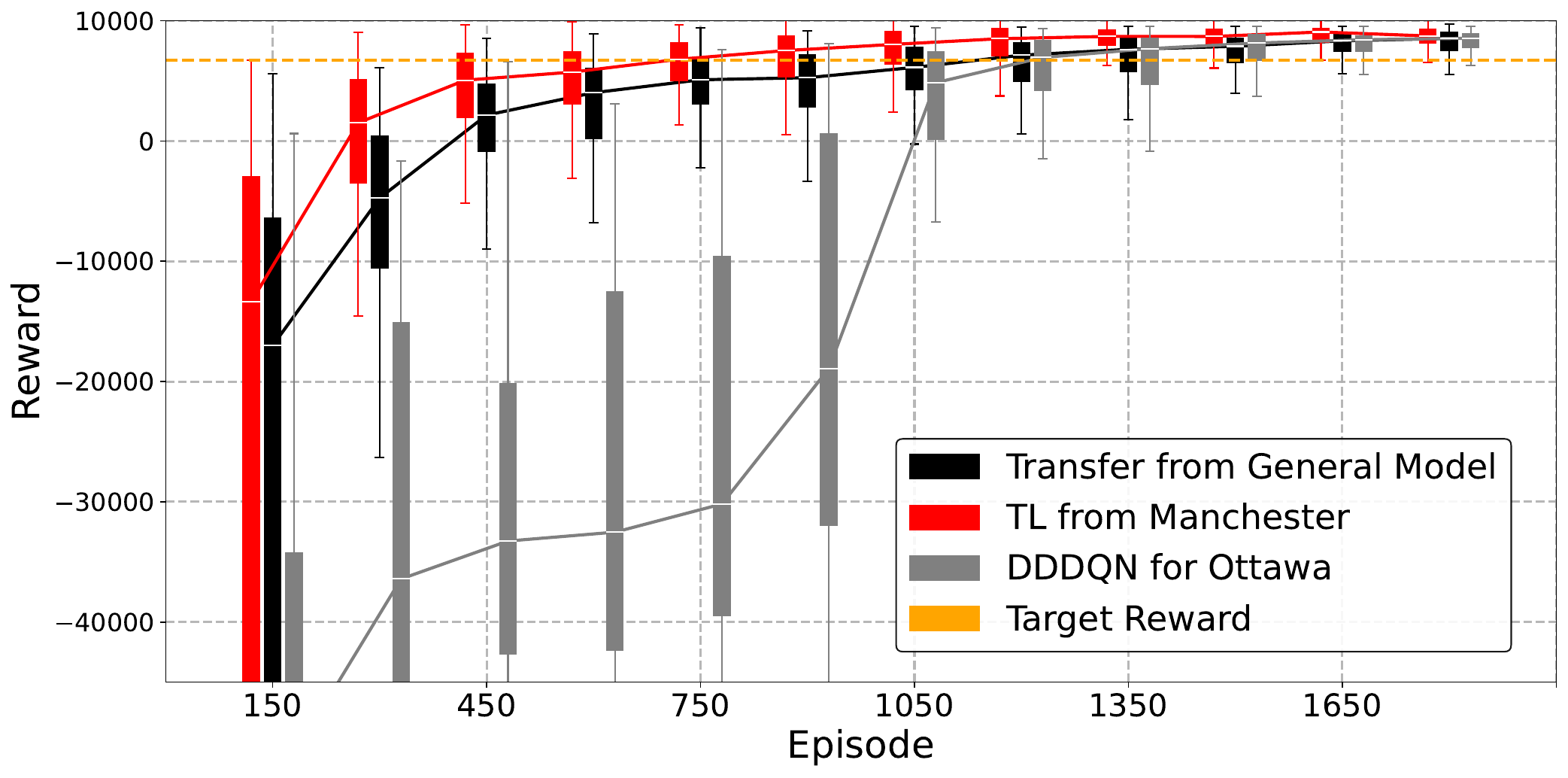}
    \caption{Performance comparison of \ac{TL} models for Ottawa.}
    \label{fig:transfer_ottawa}
\end{figure}

These results shows significant convergence improvement, Manchester outperformed Ottawa by 44.8\%, Birmingham by 34.3\%, and even York by 26.9\%. Figure~\ref{fig:transfer_ottawa} further highlights performance trajectories during training. The model from Manchester not only achieved the fastest convergence but also maintained higher and more stable rewards.

\subsubsection{Model Selection and Transfer Learning to Beijing}

\begin{figure}[h]
\centering
\begin{subfigure}[t]{0.8\columnwidth}
    \centering
    \includegraphics[width=\linewidth]{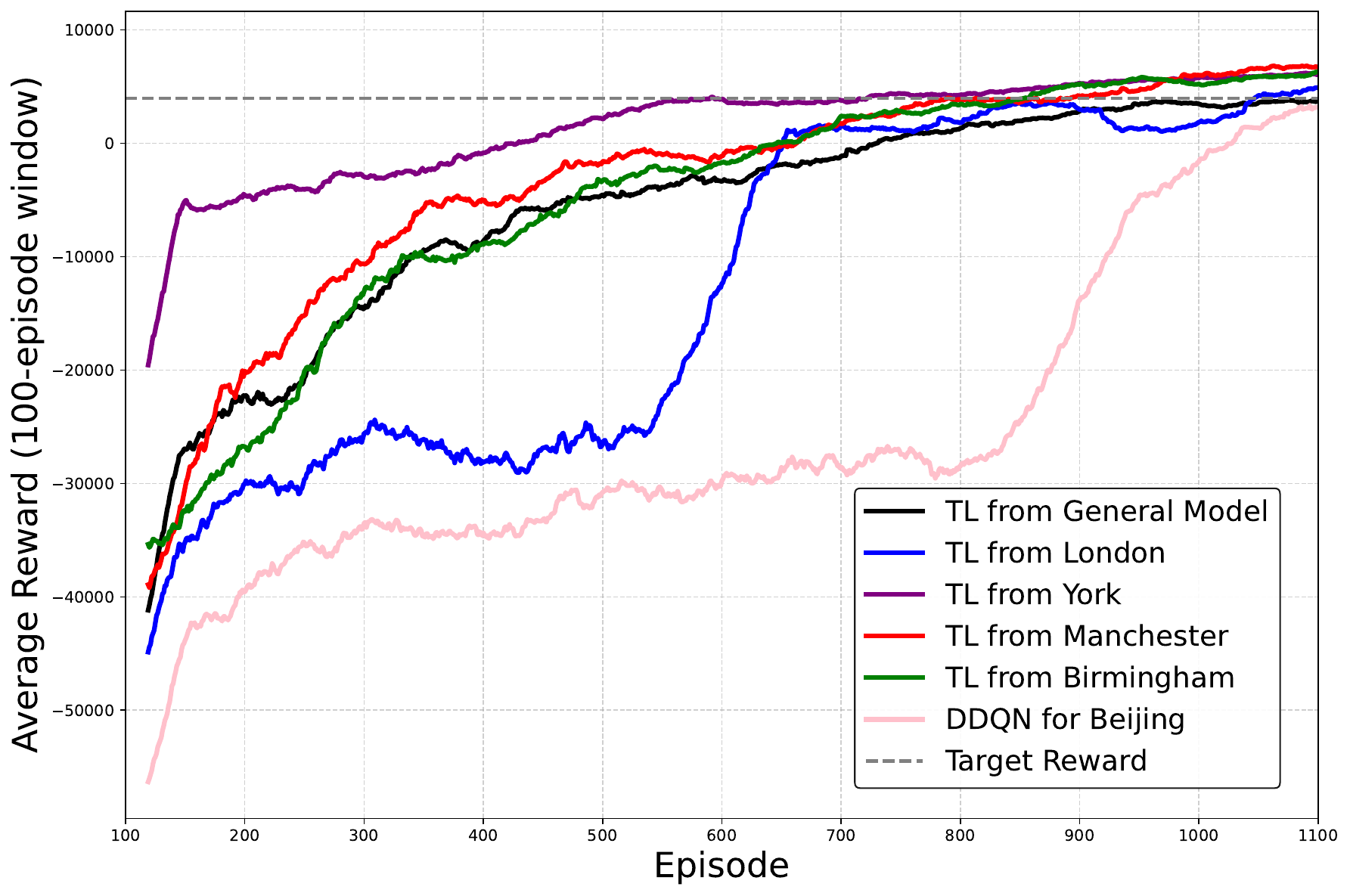}
    \caption{Full training reward curves for all compared models in the Beijing environment.}
    \label{fig:beijing_full_reward}
\end{subfigure}
\vspace{0.5em}
\begin{subfigure}[t]{0.8\columnwidth}
    \centering
    \includegraphics[width=\linewidth]{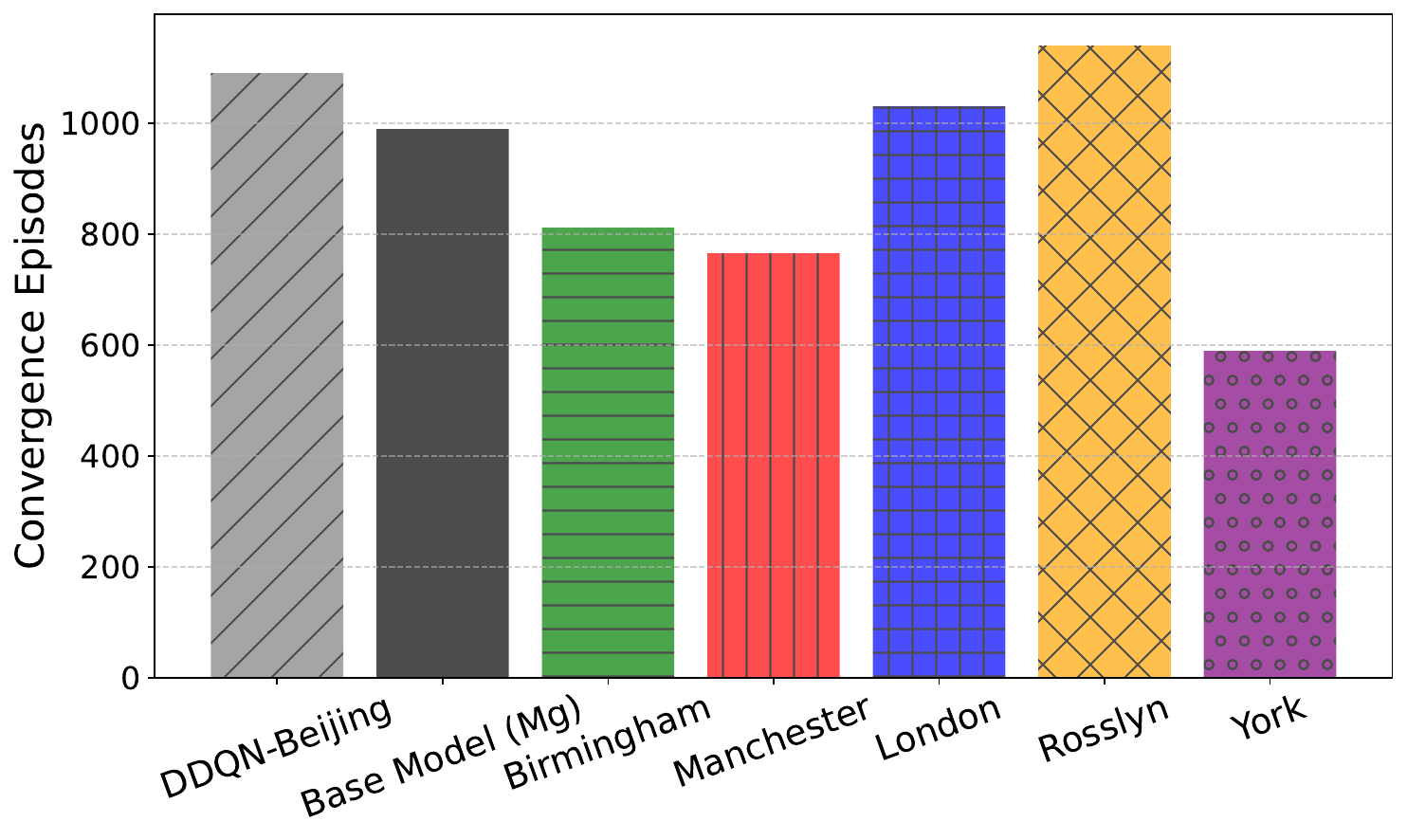}
    \caption{Performance comparison of models from the library in Beijing.}
    \label{fig:comparebeijingbar}
\end{subfigure}
\caption{Beijing environment: complete training reward curves and convergence performance comparison across different source modele.}
\label{fig:comparebeijing}
\end{figure}

\begin{table*}[h]
\centering
\caption{Performance Summary of Convergence with Statistical Variability}
\label{tab:convergence_three_methods}
\begin{tabular}{l l c c c}
\hline
\textbf{Environment} & \textbf{Method} & \textbf{Mean Episodes} & \textbf{Std. Dev. (\%)} & \textbf{Improvement vs. DDQN} \\
\hline
\multirow{3}{*}{York}
 & DDQN (Scratch) & 1380 & $\pm$6.5\% & -- \\
 & Model Selection  & 940 & $\pm$4.9\% & 31.9\% \\
 & MG ((Best model)) & 880 & $\pm$5.4\% & 36.2\% \\
\hline
\multirow{3}{*}{Ottawa}
 & DDQN (Scratch) & 1340 & $\pm$6.2\% & -- \\
 & Model Selection (Best model) & 750 & $\pm$4.6\% & 44.0\% \\
 & MG (Fallback) & 1180 & $\pm$5.1\% & 11.9\% \\
\hline
\multirow{3}{*}{Beijing}
 & DDQN (Scratch) & 1090 & $\pm$6.0\% & -- \\
 & Model Selection (Best model) & 590 & $\pm$4.4\% & 45.9\% \\
 & MG (Fallback) & 990 & $\pm$5.0\% & 9.2\% \\
\hline
\end{tabular}
\end{table*}

For the \ac{TL} evaluation in the Beijing environment, similarity scores (Table~\ref{tab:similarity}) were used to identify appropriate source models from the environment library. Among all cities, York achieved the highest similarity score (0.837), followed by Ottawa (0.703), Birmingham (0.641), and Manchester (0.628). London (0.461) and Rosslyn (0.460) had the lowest similarity scores, indicating weaker suitability for direct transfer.

Figure~\ref{fig:comparebeijing} presents the convergence episode counts of various models when transferred to Beijing. The model transferred from York achieved the fastest convergence at 590 episodes, clearly outperforming all other models. Manchester and Birmingham followed at 765 and 812 episodes, respectively. The general base model \( M_G \), trained on a synthetic average environment, required 990 episodes, while \ac{DDQN} trained from scratch in Beijing converged after 1090 episodes. In contrast, models transferred from London and Rosslyn required 1030 and 1140 episodes, respectively, indicating poor adaptability.

\begin{figure}[h!]
    \centering
    \includegraphics[width=1\columnwidth]{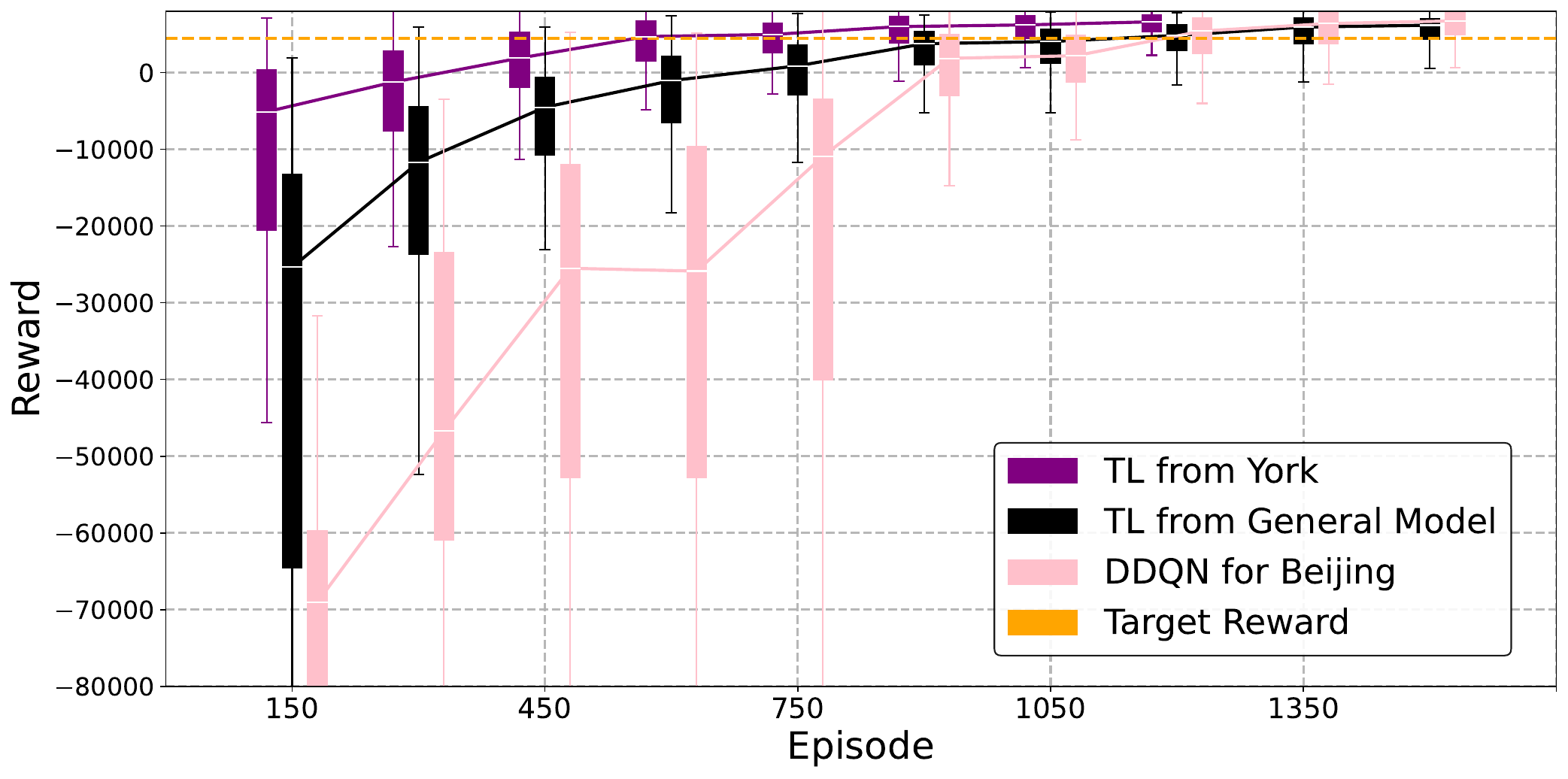}
    \caption{Performance comparison of \ac{TL} models for Beijing.}
    \label{fig:transferbeijing}
\end{figure}

These results highlight the benefits of selecting source models based on environment similarity. York outperformed \ac{DDQN} by 45.9\%, and even the base model \( M_G \) by 40.4\%. In contrast, London and Rosslyn, which differ more significantly from Beijing, failed to provide efficient knowledge transfer. Figure~\ref{fig:transferbeijing} further supports these trends by showing that the model from York achieved the highest reward growth rate and maintained consistent performance throughout training. Conversely, the models transferred from Rosslyn and London demonstrated erratic learning behavior and poor reward trajectories, consistent with their low similarity scores. Importantly, these results align well with real world geographical expectations. The York map was derived from a residential area featuring predominantly low-rise buildings, while the Beijing map was extracted from the Palace Museum (Forbidden City), which is a large, open area filled with traditional, low-height structures. This environmental similarity facilitates more effective transfer. In contrast, the London map represents a dense city center with tall, closely packed buildings, making it structurally very different from Beijing and therefore less suitable for knowledge transfer. 

A critical observation is the repeated underperformance of model from Rosslyn across all target environments. The small geographical scale and limited complexity of the Rosslyn environment provide inadequate variability for robust learning, making it a poor candidate for generalization. The chosen environment should provide a variety of spatial characteristics and sufficient complexity while not being too small.

Table~\ref{tab:convergence_three_methods} summarizes the convergence performance of three representative approaches,\ac{DDQN} training from scratch, the similarity-based model selection strategy for best transfer model, and the fallback model \(M_G\),by reporting the mean number of episodes required to reach convergence and the corresponding variability across multiple independent runs.

\subsection{Consequences of non performative model selection}
In this section, we evaluate the drawbacks of selecting a non appropriate model. As shown in Fig.~7, Fig.~8, and Fig.~10, transferring a model from a poorly matched or dissimilar source environment leads to slower convergence, increased performance variability, and in some cases inferior performance compared to training from scratch. In Environment~York city, \ac{TL} from Birmingham and \ac{DDQN} training lagged significantly, which differs from the candidate source environments in terms of building distribution, base station layout, and target location, the transferred models exhibit reduced stability and learning efficiency. These trends validate the positive correlation between architectural similarity and transfer success while also exposing exceptions such as \ac{TL} from Rosslyn and \ac{DDQN} training in Ottawa lagged significantly, it shows that despite the moderate similarity scores, the small map size limits the generalizability of the model.

\subsection{Continuous Learning of the fallback model}

\begin{figure}[t]
    \centering
    \includegraphics[width=1\columnwidth]{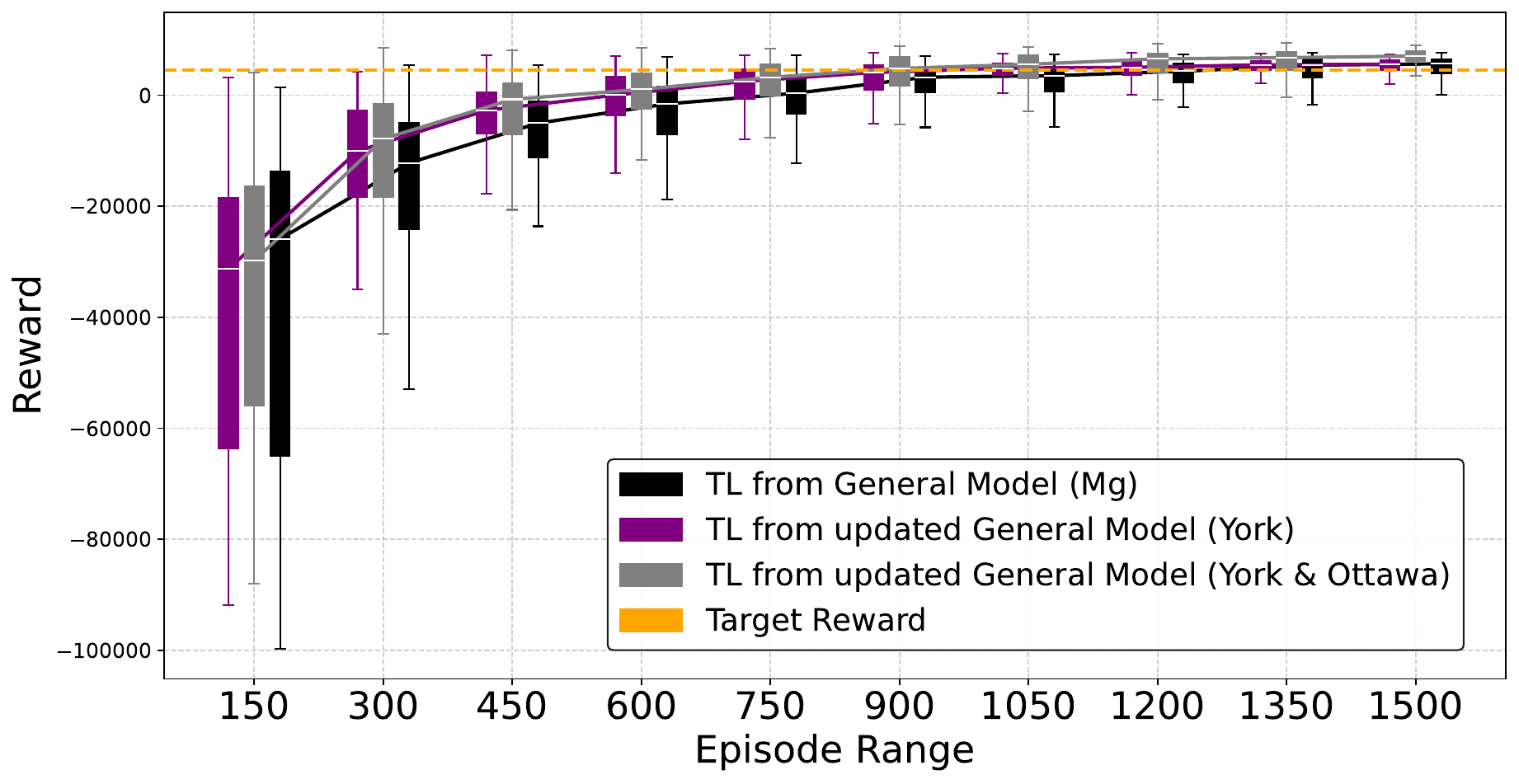}
    \caption{Performance comparison of the initial \(M_G\) and updated \(M_G\) after CL.}
    \label{fig:FL}
\end{figure}

Next, we evaluate the effectiveness of using \ac{CL} to update the generalized model \(M_G\). We base our analysis on its performance in the Beijing environment. As shown in Figure~\ref{fig:FL}, three versions of the general model are compared: the original model \(M_G\) (black), the model updated via \ac{CL} with data from York (\(M_G^1\)), and the model further updated with York and then Ottawa (\(M_G^2\)).

The results indicate that both \(M_G^1\) and \(M_G^2\) exhibit modest improvements in performance compared to the original \(M_G\), demonstrating the benefit of incorporating knowledge from past deployments. Interestingly, \(M_G^1\) and \(M_G^2\) perform similarly, suggesting that for transfer to Beijing, the knowledge gained from York contributed more significantly than that from Ottawa. This aligns with the earlier similarity analysis, where York showed the highest resemblance to Beijing among all source environments.

Although the \ac{CL} updated fallback models do not outperform the best model selected through the model selection mechanism, they still play a vital role in the broader system. Specifically, these results emphasize the importance of continuously updating the fallback model \(M_G\) with data from diverse environments. While \ac{CL} updated models may not always provide the highest performance in specific scenarios, they ensure that \(M_G\) remains a robust and adaptive fallback option.

In this context, the primary function of \ac{CL} is not to make \(M_G\) the best performing model in the short term, but rather to ensure its long-term adaptability and reliability. As the model accumulates knowledge from a growing number of heterogeneous environments, its ability to generalize and support new deployments improves over time. This highlights the significance of \ac{CL} in sustaining the evolving capability of \(M_G\), making it a resilient backbone of the \ac{UAV} trajectory optimization system.

\section{Conclusion}

In this paper, we have presented a \ac{UAV} trajectory optimization framework integrated with the \ac{O-RAN} architecture, leveraging \ac{TL} and \ac{CL} to enhance adaptability in new environments. By utilizing a structured model library and real-time trajectory adjustments, the system enables \acp{UAV} to dynamically respond to changing network conditions while minimizing training overhead. We have used \ac{CL} to ensures continuous model refinement, improving \ac{UAV} trajectory planning across diverse urban scenarios. Simulation results demonstrate that the proposed \ac{TL} with model selection approach reduces convergence time by 44\% to 56\% compared to retraining from scratch, and up to 40\% compared to traditional \ac{TL} without model selection. These gains highlight the effectiveness of intelligent model selection in accelerating adaptation while maintaining performance. The proposed approach enhances network coverage, scalability, and deployment efficiency, making it well-suited for \ac{6G} \ac{O-RAN} networks. Future work will explore further enhancements in model selection strategies and energy efficient \ac{UAV} trajectory planning.

\input{acro}

\bibliographystyle{IEEEtran}
\bibliography{references.bib}

\end{document}

%% file: acro.tex
\begin{acronym} 
\acro{5G}{Fifth Generation}
\acro{6G}{Sixth Generation}
\acro{AI}{Artificial Intelligent }
\acro{ACO}{Ant Colony Optimization}
\acro{ANN}{Artificial Neural Network}

\acro{BB}{Base Band}
\acro{BBU}{Base Band Unit}
\acro{BER}{Bit Error Rate}
\acro{BS}{base station}
\acro{BW}{bandwidth}

\acro{CAPEX}{Capital Expenditure}
\acro{CoMP}{Coordinated Multipoint}
\acro{CR}{Cognitive Radio}
\acro{CRLB}{Cramer-Rao Lower Bound}
\acro{CTL}{Continuous Transfer Learning}
\acro{CL}{continual learning} 
\acro{C-RAN}{cloud radio access network}

\acro{D2D}{Device-to-Device}
\acro{DAC}{Digital-to-Analog Converter}
\acro{DAS}{Distributed Antenna Systems}
\acro{DBA}{Dynamic Bandwidth Allocation}
\acro{DL}{deep learning}
\acro{DSA}{Dynamic Spectrum Access}
\acro{DQL}{deep Q learning}
\acro{DRL}{deep reinforcement learning}
\acro{DQN}{deep Q-network}
\acro{DDQN}{double deep Q network}
\acro{DDPG}{deep deterministic policy gradient}

\acro{FBMC}{Filterbank Multicarrier}
\acro{FEC}{Forward Error Correction}
\acro{FFR}{Fractional Frequency Reuse}
\acro{FL}{federated learning}
\acro{FSO}{Free Space Optics}
\acro{FANET}{Flying ad-hoc network}

\acro{GA}{Genetic Algorithms}
\acro{GAN}{Generative Adversarial Networks}
\acro{GMMs}{Gaussian mixture models}

\acro{HAP}{High Altitude Platform}
\acro{HL}{Higher Layer}
\acro{HARQ}{Hybrid-Automatic Repeat Request}
\acro{HCA}{Hierarchical Cluster Analysis}
\acro{HO}{Handover}
\acro{HCP}{heterogeneous computing platform}

\acro{IoT}{Internet of Things}
\acro{ISAC}{Integrated Sensing and Communication}

\acro{KNN}{k-nearest neighbors} 
\acro{KPIs}{Key Performance Indicators}

\acro{LAN}{Local Area Network}
\acro{LAP}{Low Altitude Platform}
\acro{LL}{Lower Layer}
\acro{LoS}{Line of Sight}
\acro{LTE}{Long Term Evolution}
\acro{LTE-A}{Long Term Evolution Advanced}

\acro{MAC}{Medium Access Control}
\acro{MAP}{Medium Altitude Platform}
\acro{MDP}{Markov Decision Process}
\acro{ML}{Machine Learning}
\acro{MME}{Mobility Management Entity}
\acro{mmWave}{millimeter Wave}
\acro{MIMO}{Multiple Input Multiple Output}

\acro{NFP}{Network Flying Platform}
\acro{NFPs}{Network Flying Platforms}
\acro{NLoS}{Non-Line of Sight}
\acro{Non-RT RIC}{Non-Real Time Radio Intelligent Controller}
\acro{Near-RT RIC}{Near Real-Time Radio Intelligent Controller}

\acro{O-CU}{open central unit}
\acro{O-RU}{open radio unit}
\acro{O-DU}{open distributed unit}
\acro{O-RAN}{Open Radio Access Network}
\acro{OFDM}{Orthogonal Frequency Division Multiplexing}
\acro{OSA}{Opportunistic Spectrum Access}
\acro{O-RAN}{Open Radio Access Network}
\acro{OMC}{O-RAN Management and Control}
\acro{OSM}{OpenStreetMap}

\acro{PID}{Proportional–integral–derivative}
\acro{PAM}{Pulse Amplitude Modulation}
\acro{PAPR}{Peak-to-Average Power Ratio}
\acro{PGW}{Packet Gateway}
\acro{PHY}{physical layer}
\acro{PSO}{Particle Swarm Optimization}
\acro{PU}{Primary User}

\acro{QAM}{Quadrature Amplitude Modulation}
\acro{QoE}{Quality of Experience}
\acro{QoS}{Quality of Service}
\acro{QPSK}{Quadrature Phase Shift Keying}

\acro{RU}{radio unit}
\acro{RIC}{Radio Intelligent Controller}
\acro{RAN}{Radio Access Network}
\acro{RMSE}{Root Mean Squared Error}
\acro{RN}{Remote Node}
\acro{RRH}{Remote Radio Head}
\acro{RRC}{Radio Resource Control}
\acro{RRU}{Remote Radio Unit}
\acro{RSS}{Received Signal Strength}
\acro{RF}{Radio Frequency}
\acro{RIS}{Reconfigurable Intelligent Surface}
\acro{RL}{reinforcement learning}

\acro{SAR}{synthetic-aperture radar}
\acro{SU}{Secondary User}
\acro{SCBS}{Small Cell Base Station}
\acro{SDN}{Software Defined Network}
\acro{SNR}{Signal-to-Noise Ratio}
\acro{SINR}{signal-to-interference-plus-noise ratio}
\acro{SON}{Self-organising Network}
\acro{SVM}{Support Vector Machine}

\acro{TDD}{Time Division Duplex}
\acro{TD-LTE}{Time Division LTE}
\acro{TDM}{Time Division Multiplexing}
\acro{TDMA}{Time Division Multiple Access}
\acro{TL}{transfer learning}

\acro{UE}{User Equipment}
\acro{ULA}{Uniform Linear Array}
\acro{UAV}{unmanned aerial vehicle}
\acro{USRP}{Universal Software Radio Platform}

\acro{XAI}{Explainable AI}
 
\end{acronym}